\documentclass[showpacs,aps,prd,nofootinbib,floatfix,amsmath,amssymb]{revtex4}
\usepackage{graphicx}
\usepackage[utf8x]{inputenc}
\usepackage{color}
\usepackage{subfig}
\usepackage{multirow}
\usepackage{slashed}

\begin{document}

%\preprint{IFT-P.xx/2009}
%\preprint{ArXiv:yymm.nnnn}
\title{Lepton flavor violating processes in the minimal 3-3-1 model with singlet sterile
neutrinos}

% \altaffiliation[Also at ]{Physics Department, XYZ University.}%

\author{
A. C. B. Machado
}%
\email{ana@ift.unesp.br}
\affiliation{
Instituto  de F\'\i sica Te\'orica--Universidade Estadual Paulista \\
R. Dr. Bento Teobaldo Ferraz 271, Barra Funda\\ S\~ao Paulo - SP, 01140-070,
Brazil
}

\author{
J. Monta\~no
}%
\email{montano@ift.unesp.br}
\affiliation{
Instituto  de F\'\i sica Te\'orica--Universidade Estadual Paulista \\
R. Dr. Bento Teobaldo Ferraz 271, Barra Funda\\ S\~ao Paulo - SP, 01140-070,
Brazil
}

\author{V. Pleitez}%
\email{vicente@ift.unesp.br}
\affiliation{
Instituto  de F\'\i sica Te\'orica--Universidade Estadual Paulista \\
R. Dr. Bento Teobaldo Ferraz 271, Barra Funda\\ S\~ao Paulo - SP, 01140-070,
Brazil
}

\date{11/21/2016}
%07/11/15% It is always \today, today,

\begin{abstract}

\

We consider the minimal 3-3-1 model with three sterile neutrinos transforming as singlet under the $SU(3)_L\otimes U(1)_X$ symmetry. This model, with or without sterile neutrinos, predicts flavor violating interactions in both quark and lepton sectors, since all the charged fermions mass matrices can not be assumed diagonal in any case.
Here we accommodate the lepton masses and the Pontecorvo-Maki-Nakawaga-Sakata matrix at the same time, and as consequence the Yukawa couplings and the unitary matrices which diagonalize the mass matrices are not free parameters anymore.
We study some phenomenological consequences, i.e., $l_i\to l_jl_k \bar{l}_k$ and $l_i\to l_j\gamma$ which are induced by neutral and doubly charged particles present in the model. In particular we find that if the decay $\mu\to ee\bar{e}$ is observed in the future, the only particle in the model that could explain this decay is the doubly charged vector bilepton.

\end{abstract}


\maketitle

\section{Introduction}
\label{sec:intro}

It is well known that models with 3-3-1 gauge symmetries predict flavor violating interactions in both quark and lepton sectors, since all the charged fermions mass matrices can not be assumed diagonal in any case \cite{Boucenna:2015zwa}.
On the other hand, it has been confirmed that neutrinos are massive particles and that there is mixing in the leptonic charged current \cite{Agashe:2014kda}. In the context of the standard model (SM), the massive neutrinos imply that the charged currents are non-diagonal and they are parametrized by the Pontecorvo-Maki-Nakagawa-Sakata (PMNS) matrix, which has been measured in several neutrino oscillation experiments~\cite{Agashe:2014kda,GonzalezGarcia:2012sz}. Nevertheless, in the SM there is no mechanism for generating neutrino masses neither the PMNS matrix. Hence, it is necessary to search extensions of the SM that implement a mechanism for solving both issues, keeping at the same time compatibility with the data of lepton flavor violating (LFV) processes.

Here we will implement a mechanism for generating lepton masses and the PMNS matrix in the context of the minimal 3-3-1 (m331) model in which the left-handed lepton families transform as triplets $\Psi=(\nu_{l_i},\,l_i,\,l_i^c)^T_L\sim(\textbf{3},0)$ under the $SU(3)_L\otimes U(1)_X$ symmetry~\cite{Pisano:1991ee,Frampton:1992wt,Foot:1992rh}. We will also study the phenomenological effects of such mechanism in some LFV processes like $l_i\to l_jl_k\bar{l}_k$, where $l_i=\mu,\tau$, $l_{j,k}=e,\mu$, $l_i\to l_j\gamma$ and $h^0\to l_i\bar{l}_j$.

Moreover, we will work in the context of the non-trivial SM limit of our model obtained when we impose $\rho_0=1$, where  $\rho_0=m^2_W/c^2_Wm^2_{Z_1}$, being $m_{Z_1}$ and $m_W$ the masses of the respective
lightest $Z_1$ neutral and $W$ charged vector bosons, corresponding to the $Z$ and $W$ gauge bosons in the SM. Under this condition, all couplings of the known fermions to $Z_1$ are the same as the respective couplings of these fermions to $Z$ in the SM, and the exotic quarks couplings  to $Z_1$ and $Z_2$ depend only on the electroweak angle $\theta_W$. In the m331 model, in which the lepton sector includes only the known leptons, this condition fixes the vacuum expectation values (VEVs) in terms of $\theta_W$~\cite{Dias:2006ns},
\begin{equation}
v^2_\rho=\frac{1-4s^2_W}{2c^2_W}v^2_W,\quad v^2_\eta+2v^2_{s_2}=\frac{1+2s^2_W}{2c^2_W}v^2_W,
\label{solution}
\end{equation}
only $v_\chi$ remains free. Here $v^2_W$ is related to the standard electroweak scale, and $v_{s_2}$ is the VEV of the neutral component of the sextet which contributes to the charged lepton masses, $v_\eta$ and $v_\rho$ are the VEVs of the triplets that contribute to the quark masses, but $\eta$ also may contribute to the charged lepton masses.  Analysis of the quark sector assuming (\ref{solution}) was done in Ref.~\cite{Machado:2013jca}, there was assumed that the VEVs of the triplets $\eta$ and $\rho$ satisfy $v_\eta^2+v_\rho^2\approx (246\ \textrm{GeV})^2$. Hence, $v_{s_2}$ was considered negligible and for this reason the sextet is not enough, as we will show below, to give the correct mass to the charged leptons using only renormalizable Yukawa interactions. Notice that this fixes three otherwise almost arbitrary VEVs (in GeV): $v_\eta\sim 240, v_\rho\sim 54$, and $2v_{s_2}^2< 246^2-v_\rho^2-v_\eta^2$. However, since the degrees of freedom of the sextet are still there and may be heavy, they can induce the  dimension five effective operator given in Ref.~\cite{DeConto:2015eia}. For these reasons it is necessary to add to the m331 model, three sterile neutrinos. See more details in Sec.~\ref{sec:model}.

The m331 model has doubly charged vector and scalar bileptons, generically denoted by $X^{--}$, with interactions that violate the individual lepton flavor number by single or two units, for instance $X^{--}\to \mu^-e^-$ and $X^{--}\to e^-e^-$,  respectively. Hence, in this models the decays $\tau\to \mu\mu\bar{\mu},ee\bar{e}$ (the three leptons may be all different) and $\mu\to ee\bar{e}, e \gamma,...$ are allowed and can be used to constraint the masses of the bileptons once the unitary matrices, $V^l_{L,R}$  and $V^\nu_L$, needed to diagonalize the lepton matrices, are fixed. In fact, these processes are prediction of this sort of model~\cite{Boucenna:2015zwa}.

These flavor violation processes in the m331 model were considered many years ago by Liu and Ng~\cite{Liu:1993gy}. However, at that time almost nothing was known about the lepton mixing and neutrino masses. Here we will take into account this new information and also consider the effects of the neutral Higgs bosons which in this model have non-diagonal interactions in the flavor space. We stress that even if neutrinos were massless (at tree level) the lepton flavor number is not conserved in this model. Here we will not consider neither $C\!P$ violation, however see~\cite{DeConto:2014fza}.

Generally, in models with FCNC the unitary
matrices, $V^l_{L,R}$ and $V^\nu_{L}$,that are needed to diagonalize the mass matrix in each charged
sector, survive in different places of the Lagrangian when it is written
in terms of the mass eigenstates fields.
For instance, in the present model, besides the $V_{PMNS}=V^{l\dagger}_LV^\nu_L$ matrix in the lepton sector we have interactions with the doubly charged vector bilepton field that involve $V_U=(V_R^{l})^TV^l_L$, and with doubly charged scalars with $K_R=(V_R^{l})^\dagger G^sV^{l *}_R$ where
$G^s$ is a symmetric matrix of Yukawa couplings.
For this reason, we first fit all of these unitary
matrices by getting the known masses and mixing matrices in the
interactions with $W_\mu^\pm$ and only then study the phenomenological
consequences constraining the masses of the extra particles in the
model.

In the absence of a compelling model new physics effects can be parametrized by using
effective interactions in which all operators of a given dimension are classified. For instance, effective operators which violate flavor, baryon and lepton numbers
can be  used  in a model independent  way. These operators depends on the unitary matrices above and on an energy scale, $\Lambda$, which characterizing the new physics and  is larger than the electroweak scale~\cite{Buchmuller:1985jz,Leung:1984ni,Chang:2005wu,Dorsner:2015mja}. However, in the present case, we have a well behaved model and all new interactions are formulated using only renormalizable interactions, since even the effective interactions used for generating the charged lepton masses arising as a consequence of fundamental Yukawa interactions of leptons with the sextet.
Moreover, even if effective operators are used to fit the unitary matrices and the $\Lambda$ scale, eventually we have to verify if these unitary matrix entries obtained in this way, do correctly diagonalize the respective fermion masses.

Our main result, using the strategy discussed above, i.e., first fit all unitary matrices, is that the strongest constraint on the doubly charged vector bilepton $U^{++}_\mu$, comes from $\mu\to ee\bar{e}$. We will show that unlike processes in the quark sector~\cite{Machado:2013jca,Correia:2015tra} in leptonic processes the scalar contributions are supressed. This sort of models predict also flavor changing neutral currents (FCNC) in the scalar sector, thus we consider also $h^0\to l_i\bar{l}_j$, at tree level, in a situation in which there is no more free parameters in the Yukawa coupling since all of them are already fixed when we obtain the matrices $V^l_{L,R},V^\nu_L$, the PMNS mixing matrix and lepton masses.

The structure of the article is as follows, in Sec.~\ref{sec:model} we review briefly the m331 model \textsl{plus} three sterile (under the 3-3-1 gauge symmetry) neutrinos. In Sec.~\ref{sec:chargedleptons} we present the mass matrices of the lepton sector, as well a numerical parametrization for the $V^l_{L,R},V^{\nu}_L$ matrices that will be used in the present study. The interactions with the doubly charged vector bosons are presented in Sec.~\ref{subsec:2vector}, and the interactions with the doubly charged and neutral scalars are presented in Sec.~\ref{subsec:2scalars}. In Sec.~\ref{sec:pheno} we present the phenomenological consequences of our analysis. The last section \ref{sec:con} is devoted to our conclusions. We comment three other numerical parametrization for the matrices $V^l_{L,R}$ in the Appendix~\ref{sec:matrices}, this is done just to show how the constraints on $m_{U^{++}}$ depend on those matrices. Finally, in Appendix~\ref{sec:amplitudes} we give samples of the amplitudes of the vector doubly charged bilepton contribution.

\section{The m331 with singlet right-handed neutrinos}
\label{sec:model}

In the m331 model the scalar sector is conformed by  $\eta=(\eta^0,\eta^{-}_1,\eta^+_2)^T\sim({\bf3},0)$,
$\rho=(\rho^+,\rho^0,\rho^{++})^T\sim({\bf3},1)$,
$\chi=(\chi^-,\chi^{--},\chi^0)^T\sim({\bf3},-1)$, and the sextet  $(\textbf{6},0)$
\begin{equation}
S=\left(
\begin{array}{ccc}
s^0_1& \frac{s^-_1}{\sqrt2} & \frac{s^+_2}{\sqrt2}\\
\frac{s^-_1}{\sqrt2}& S^{--}_1&\frac{s^0_2}{\sqrt2}\\
\frac{s^+_2}{\sqrt2}&\frac{s^0_2}{\sqrt2}&S^{++}_2
\end{array}
\right),
\label{sextet1}
\end{equation}
where the numbers between parentheses denote the transformation properties under $SU(3)_L\otimes U(1)_X$, respectively.
In this model there are only left-handed neutrinos and they are naturally Majorana particles. The sextet $(\textbf{6},0)$, in principle, gives mass to both the charged leptons and neutrinos. The Yukawa interactions are given by
\begin{equation}
-\mathcal{L}_l=\epsilon_{ijk}\overline{(\Psi)^c}_{ia}G^\eta_{ab}\Psi_{bj}\eta_k +
\overline{(\Psi)^c}_{ia}G^S_{ab}\Psi_{bj}S^*+H.c.,
\label{m331leptons1}
\end{equation}
and  we obtain the following mass matrices in the lepton sector
\begin{equation}
M^l=\frac{v_\eta}{\sqrt2}G^\eta+\frac{v_{s_2}}{2}G^S,\quad
M^\nu=\frac{v_{s_1}}{\sqrt2}G^S.
\label{m331mass1}
\end{equation}
Note that the matrix for the charged leptons has two contributions, one from a symmetric matrix $G^S$  and other from an antisymmetric matrix $G^\eta$. An $3\times 3$ antisymmetric matrix has the eigenvalues $\{0, -m, m \}$, therefore, to adjust the mass of charged leptons, the largest contribution would have to come from the symmetric matrix, plus a minor contribution from the antisymmetric matrix. This could be possible if $v_\eta \ll v_{s_2}$, however, as we said above that $v_\eta\sim 240$ GeV, hence $v_\eta \gg v_{s_2}$, condition which is incompatible with the SM limit discussed in the Introduction.
At first sight this problem could be avoided if the interaction with $\eta$ is forbidden.
However, if we avoid this interaction, the mass matrix of the charged lepton is proportional to the neutrino mass matrix and, for this reason, both mass matrices are diagonalized by the same unitary matrix and it is not possible to obtain a realistic PMNS matrix defined as $V_{PMNS}=V^{l\dagger}_LV^\nu_L$ since in this case $V_{PMNS}=\textbf{1}$.

One possibility for generating a realistic lepton mass spectra could be generated by radiative corrections if there are interactions in the scalar sector that violates explicitly the conservation of $L$. The assignment of the total lepton number $L$ is:
\begin{equation}
L(\eta^-_2,\chi^-,\chi^{--},\rho^{--},V^-_\mu,U^{--}_\mu,s^-_1,s^0_1,S^{--}_1,S^{++}_2)=+2,
\label{ln}
\end{equation}
only $\eta^-_1,\eta^0,\rho^+,\rho^0,\chi^0,s^+_2,s^0_2$ carry $L=0$.

Another way, that we will use here, is introducing right-handed neutrinos, singlet of $SU(3)_L\otimes U(1)_X$ implementing a type-I seesaw mechanism.
The only source of total lepton number violation is the Majorana mass term for the right-handed neutrinos.
In this case the lepton masses are generated by the triplets $\eta,\rho$ and $\chi$ and a dimension five operator induced by the heavy sextet  as in Ref.~\cite{DeConto:2015eia}. Hence, the Yukawa interactions are
\begin{eqnarray}
-\mathcal{L}^{lep}_Y &=&-\frac{1}{2}\epsilon_{ijk}\,\overline{(\Psi_{ia})^c}G^\eta_{ab} \Psi_{jb}\eta_k+\frac{1}{2\Lambda_s}\,
\overline{(\Psi_{a})^c} G^s_{ab} (\chi^*\rho^\dagger+ \rho^*\chi^\dagger)\Psi_{b} \nonumber \\ &+&
\overline{(\Psi_{aL})}G^\nu_{ab}\nu_{aR}\eta+ \overline{(\nu_{aR})^c}M_{ab}\nu_{bR}+H.c.
\label{effective1}
\end{eqnarray}
where $\Lambda_s$ is a mass scale generated by the effective interactions induced by the heavy scalar.
It means that FCNC processes in the lepton and quark sector are predictions of this model. The scalar contributions can not, in general, be neglected anymore. At least this is the case in the quark sector~\cite{Machado:2013jca,Correia:2015tra}.

\section{The lepton mass matrices}
\label{sec:chargedleptons}

From (\ref{effective1}) the lepton mass matrices are given by
\begin{equation}\label{mmassa3}
M^\nu_{ab}\approx -\frac{v^2_\eta}{2}G^\nu\frac{\bar{M}}{M_R}G^{\nu T},\quad
M^l_{ab}=G^\eta_{ab}\frac{v_\eta}{\sqrt2}+\frac{1}{\Lambda_s}G^s_{ab}v_\rho v_\chi.
\end{equation}
where $M_R$ is a mass scale generated by the violation of the lepton number $L$. An interesting possibility is when $M$ is diagonal and $M_3=M_R\gg M_1,M_2$. Defining  $r_i\equiv M_i/M_R,\;i=1,2$, we have $M=M_R\textrm{diag}(r_1\,r_2,1)$ that  $M^{-1}=(1/M_R)\bar{M}$ where $\bar{M}=\textrm{diag}(\bar{r}_{1},\bar{r}_{2},1)$ and $\bar{r}_i\equiv M_R/M_i$. For the sake of simplicity we will assume that all right-handed neutrinos are mass degenerated and $M^{-1}=(1/M_R)\textbf{1}$. Moreover, in the following we assume  $v_\chi\approx \Lambda_s$ and, as in Ref.~\cite{Machado:2013jca}, $v_\rho\sim 54,v_\eta\sim 240$ GeV, and $M_R\sim 1$ TeV.

The mass matrices in the charged sector $M^l$ and in the neutrino sector $M^\nu$ are diagonalized as  $\hat{M}^\nu=V^{\nu T}_LM^\nu V^\nu_L$ and $\hat{M}^l=V^{l\dagger}_L M^l V^l_R$, where $\hat{M}^\nu = diag (m_1, m_2, m_3)$ and $\hat{M}^l = diag (m_e, m_\mu, m_\tau)$. The relation between symmetry eigenstates (primed) and mass (unprimed) fields are $l^\prime_{L,R}=V^l_{L,R}l_{L,R}$
and $\nu^\prime_L=V^\nu_L \nu_L$, where
$l^\prime_{L,R}=(e^\prime,\mu^\prime,\tau^\prime)^T_{L,R}$, $l_{L,R}=(e,\mu,\tau)^T_{L,R}$  and
$\nu ^\prime_L=(\nu_e,\nu_\mu,\nu_\tau)^T_L$
and $\nu_L=(\nu_1,\nu_2,\nu_3)_L$.

We will solve simultaneously the equations
\begin{equation}
V^{l\dagger}_L M^l M^{l\dagger}V^l_L=V^{l\dagger}_R M^{l\dagger}  M^lV^l_R=(\hat{M}^l)^2 \quad (a), \quad
\hat{M}^\nu=(V^{\nu }_L)^T M^\nu V^\nu_L \quad (b),
\label{eq1}
\end{equation}
in order to obtain the matrices $V^l_{L,R}$ and $V^\nu_L$, and at the same time the PMNS matrix defined as $V_{PMNS}=V^{l\dagger}_L V^\nu_L$.
For that we must assign values to the Yukawas, $G^\eta_{ab}$, $G^s_{ab}$ and $G^\nu_{ab}$, using a computational routine in \texttt{Mathematica} program that solves simultaneously the Eqs.~(\ref{eq1}) and  $V_{PMNS}=V^{l\dagger}_L V^\nu_L$. First, let us consider the charged lepton masses in Eq.~(\ref{eq1}a).
As a result the we obtain the following values for the Yukawa couplings: the symmetric matrix with
$G^s_{11} = 5 \times 10^{-8}$, $G^s_{12} = 0.0001402$, $G^s_{13} =5 \times 10^{-8}$, $G^s_{22} = 5 \times 10^{-8}$, $G^s_{23} = 0.008$, $G^s_{33} =0.03096$  and the antisymmetric matrix $G^\eta_{12} = G^\eta_{13} = G^\eta_{23} = 10^{-5}$. With them we found $\hat{M}^{l} = \textrm{diag}(0.000511, 0.105, 1.776)$ GeV, and the diagonalization matrices are
\begin{equation}
V^l_L\approx\left(\begin{array}{ccc}
0.997501   & 0.0706425 & 0.00112211 \\
0.0683744  & -0.96923  & 0.23647 \\
-0.0177924 & 0.235802  & 0.971638 \\
\end{array}\right),\quad
V^l_R\approx\left(\begin{array}{ccc}
0.997666   & 0.0682713 & 0.000892181 \\
0.0661281  & -0.969434 & 0.236274 \\
-0.0169956 & 0.235664  & 0.971686 \\
\end{array}\right).
\label{vlr1}
\end{equation}
Next, to solve the Eqs.(\ref{eq1}b) and  $V_{PMNS}=V^{l\dagger}_L V^\nu_L$ we assume that the latter matrix, taken from Eq.~(2.1) of Ref.~\cite{GonzalezGarcia:2012sz}, can vary within the $3\sigma$ experimental error range, resulting in the following values for the Yukawas couplings $(\times 10^{-9})$: $G^\nu_{11} = 0.001475 - 0.001601$, $G^\nu _{12} = - (0.002994 - 0.000996)$, $G^\nu _{13} = -0.00452 - 0.001169$, $G^\nu _{22} = 0.008987- 0.004015$, $G^\nu _{23} = - (0.003125 - 0.00944)$, $G^\nu _{33} = 0.01833 - 0.02317$. With them we obtain
\begin{equation}
V_L^\nu\approx\left(\begin{array}{ccc}
0.85 - 0.89 & 0.42 - 0.52 & 0.11 - 0.19 \\
0.40 - 0.49 & 0.69 - 0.91 & 0.14 - 0.53
\\
0.20 - 0.23 & 0.05 - 0.51 & 0.84 - 0.97
\\
\end{array}\right),
\label{vlnus1}
\end{equation}
where we assume the normal mass hierarchy $\hat{M^\nu} = (0, \sqrt{\Delta m_{12}^2}, \sqrt{\Delta m_{23}^2})$  for the central values in PDG. Thus from $V_L^l$ in (\ref{vlr1}) and $V_L^\nu$ in (\ref{vlnus1}):
\begin{equation}
|V_{PMNS}|\approx\left(\begin{array}{ccc}
0.81-0.86 & 0.51-0.53 & 0.07-0.23 \\
0.46-0.53 & 0.54-0.82 & 0.54-0.65 \\
0.23-0.23 & 0.26-0.65 & 0.54-0.72 \\
\end{array}\right),
\label{pmns1}
\end{equation}
note that these range of values overlap those given in Ref.~\cite{GonzalezGarcia:2012sz}. If in contrast we assume the inverse mass hierarchy $\hat{M^\nu} = (\sqrt{\Delta m_{23}^2}, \sqrt{\Delta m_{12}^2}, 0)$ we have to interchange in the Eq.~(\ref{vlnus1}) the first and the third column as a consequence we will obtain a totally different Eq.~(\ref{vlr1}) in order to obtain Eq.~(\ref{pmns1}).

The phenomenology of the model depends strongly on the numerical values of the matrices in Eq.~(\ref{vlr1}), (\ref{vlnus1}) and (\ref{pmns1}). To show this, in the Appendix \ref{sec:matrices} we present other parametrizations of the matrices and we discuss in the Conclusions how the constraint on the mass of the vector bilepton $U_\mu^{++}$, coming from the decay $\mu\to ee\bar{e}$, is affected when other parametrizations are used.

\section{Interactions}
\label{sec:interactions}

The matrices $V^l_{L,R}$ in (\ref{vlr1}) will appear in the leptonic interactions. In Sec.~\ref{subsec:2vector} we consider these interactions with doubly charged vector bosons and in Sec.~\ref{subsec:2scalars} with doubly charged and neutral scalars.

\subsection{Interaction with the doubly charged vector bosons}
\label{subsec:2vector}

The interactions of leptons with the doubly charged vector boson $U^{\pm\pm}_\mu$ are obtained from the Lagrangian
\begin{equation}
\mathcal{L}^{leptons}=\frac{1}{2}\left(\overline{\Psi}\gamma^\mu \mathcal{D}_\mu\Psi +\overline{\Psi^c}\gamma^\mu \mathcal{D}^c_\mu\Psi^c \right),
\label{uu1}
\end{equation}
where the covariant derivatives for leptons $\Psi\sim(\textbf{3},0)$ are defined in the m331 model as
\begin{equation}
\mathcal {D}_\mu=\partial_\mu\textbf{1} + ig {\cal M}_\mu,\quad\mathcal{D}^c= \partial_\mu\textbf{1} - ig {\cal M}^*_\mu,
\label{dcl}
\end{equation}
with ${\cal M}_\mu\equiv \vec{W}_\mu\cdot \vec{T}$ we get
\begin{eqnarray}
{\cal{M}}_\mu= \left(\begin{array}{ccc}
W^3_\mu + \frac{1}{\sqrt 3}W^8_\mu & \sqrt{2}W_\mu^{+} &
\sqrt{2}V^-_\mu \\
\sqrt{2}W_\mu^{-} & -W^3_\mu+ \frac{1}{\sqrt 3}W^8_\mu &
\sqrt{2}U^{--}_\mu   \\
\sqrt{2}V^+_\mu  & \sqrt{2}U^{++}_\mu  & -\frac{2}{\sqrt 3}W^8_\mu
\end{array}\right),
\label{mbv}
\end{eqnarray}
where the non hermitian gauge bosons are defined as
\begin{eqnarray}
W_\mu^{+}=(W^1_\mu-iW^2_\mu)/\sqrt 2,\quad
V^+_\mu =(W^4_\mu+iW^5_\mu)/\sqrt 2, \quad
U^{++}_\mu =(W^6_\mu+iW^7_\mu)/\sqrt 2.
\label{defbc}
\end{eqnarray}

In this case the charged current coupled to the doubly charged vector boson as
\begin{align}\label{uu2}
\mathcal{L}^U =& -\frac{g}{\sqrt{2}}\left[\overline{(l^c)_{aL}}\gamma^\mu V_{Uab}l_{bL}
-\overline{(l_{aL})^c}\gamma^\mu V_{Uab}^Tl_{bR}
\right] U_\mu^{++} +H.c. \nonumber\\
    =& -\frac{g}{2\sqrt2}\,\bar{l^c_a}\gamma^\mu\left(V_{Uab}P_L-V_{Uab}^TP_R\right)l_b\,
    U^{++}_\mu+H.c.\nonumber \\
    =&- \frac{g}{2\sqrt2}\,\bar{l^c_a}\gamma^\mu\left(V_{Vab}-V_{Aab}\gamma_5\right)l_b\, U^{++}_\mu+H.c.,
\end{align}
whith $P_{R,L}=(\textbf{1}\pm\gamma_5)/2$ and $V_U \equiv V^{lT}_{R} V^l_{L}$, where $V^l_{R,L}$ given in Eq.~(\ref{vlr1}) are the matrices that diagonalize the charged lepton matrices.
The second line shows that the interactions can be split in left- and right-handed currents, as first noted by Liu and Ng~\cite{Liu:1993gy}. In the third line we have split the interaction in vector and axial vector currents via the introduction of an antisymmetric matrix $V_V=(V_U-V^T_U)/2$ and a symmetric one $V_A=(V_U+V^T_U)/2$. Notice that the diagonal family couplings are purely axial vector. However, in our calculations we will only use left-handed currents, see the discussion in the Appendix \ref{sec:amplitudes}.

Using the numerical values of the matrices $V^l_{L,R}$ in (\ref{vlr1}) we get
\begin{equation}\label{vu1}
V_U\approx\left(\begin{array}{ccc}
0.999997     & 0.00237676  & 0.000243171 \\
-0.00237672  & 0.999997    & -0.000185706 \\
-0.000243612 & 0.000185128 & 1 \\
\end{array}
\right) ,
\end{equation}
and
\begin{equation}\label{vu1-vv-va}
V_V=\left(
\begin{array}{ccc}
 0 & 0.00237674 & 0.000243391 \\
 -0.00237674 & 0 & -0.000185417 \\
 -0.000243391 & 0.000185417 & 0 \\
\end{array}
\right)
 , \
V_A=
\left(
\begin{array}{ccc}
0.999997 & 10^{-8}  & -10^{-7} \\
10^{-8}  & 0.999997 & -10^{-7} \\
-10^{-7} & -10^{-7} & 1 \\
\end{array}
\right) .
\end{equation}
Notice that this parametrization of $V_U$ is almost a diagonal matrix, and this characteristic will impact in the bilepton mass prediction derived from the $\mu\to ee\bar{e}$ experimental upper limit. This behaviour will be discussed in detail in Sec.~\ref{sec:3leptons}. Moreover, we can see from (\ref{vu1-vv-va}) that $V_V$, which is antisymmetric by construction, has its off-diagonal entries very small in this parametrization. However, we stress that unlike Refs.~\cite{Barreto:2013paa,RamirezBarreto:2011av}, we can not assume $V_V$ in the third line in (\ref{uu2}) to be exactly zero because in that case $V_U$ would be a diagonal matrix, which is not possible by the definition $V_U \equiv V^{lT}_{R} V^l_{L}$, and notice also that from $M^l$ in (\ref{mmassa3}) we can not assume such matrices to be diagonal since the beginning.

\subsection{Interactions with the doubly charged and neutral scalars}
\label{subsec:2scalars}

From Eq.~(\ref{effective1}) we obtain the interactions with the mass eigenstate leptons
with the doubly charged scalar bosons:
\begin{align}\label{ss}
\mathcal{L}^Y_{CC}=&-\frac{1}{\Lambda}\left(\frac{2v_\rho}{\sqrt2}\overline{l^c_{aR}}\,K_{Lab} \,l_{bL}\chi^{++}
    +\frac{2v_\chi}{\sqrt2}\overline{l_{aR}}\,K_{Rab}\,l^c_{bL}\rho^{--}\nonumber \right.\\
    & +\left. \frac{2v_\rho}{\sqrt2}\overline{l_{aL}}\,K^*_{Lab} \,l^c_{bR}\chi^{--}+\frac{2v_\chi}{\sqrt2}\overline{l^c_{aL}}\,K^*_{Rab}\,l_{bR}\rho^{++}\right)+H.c.\nonumber \\
    &\approx-\frac{\sqrt{2}v_\chi}{\Lambda}\left(\overline{l_{aR}}\,K_{Rab}\,l^c_{bL}\rho^{--}
    +\overline{l^c_{aL}}\,K^*_{Rab}\,l_{bR}\rho^{++} \right)
\end{align}
where in the third line only the interactions proportional to $v_\chi$ were kept, and we have defined
\begin{equation}
K_{L}=V^{lT}_L\,G^s\,V^l_L,\quad K_R=V^{l\dagger}_R\,G^s\,V^{l*}_R.
\label{defk}
\end{equation}
Since $G^s$ is a symmetric matrix, $K_L$ and $K_R$ are symmetric matrices.
For the sake of simplicity we will consider the contribution of just one of the four physical doubly charged scalars, here denoted by $Y_1^{++}$, and we assume that the other three are, heavy enough or suppressed by the matrix elements, hence $\rho^{++}\approx O_{\rho1}Y^{++}_1$ and $\chi^{++}\approx O_{\chi1}Y^{++}_1$. Using the Yukawa couplings in Sec.~\ref{sec:chargedleptons}, and the the matrices in (\ref{vlr1}) we obtain:
\begin{equation}\label{klkr}
K_L\approx
\left(
\begin{array}{ccc}
0.949736 & 0.23483  & -0.395744 \\
         & -195.657 & 0.321939  \\
         &	        & 3290.41   \\
\end{array}
\right)\times10^{-5} \ , \
K_R\approx
\left(
\begin{array}{ccc}
0.949744 & -0.232454 & 0.404881  \\
         & -195.657  & -0.324448 \\
         &           &	3290.41   \\
\end{array}
\right)\times10^{-5} \ .
\end{equation}\label{a1e11}
The charged current of interest for our phenomenology studies are
\begin{equation}
\mathcal{L}^Y_{CC}=-
\frac{2v_\chi}{\Lambda}O_{\rho1}\,
\left(\overline{l_{aR}}\,K_{Rab}\,l^c_{bL}Y^{--}_1+\overline{l^c_{aL}}\,K^*_{Rab}\,l_{bR}Y^{++}_1\right).
\label{ss2}
\end{equation}
There are also FCNC interactions of leptonse with neutral scalars and pseudo-scalars given by
\begin{eqnarray}
\mathcal{L}^h &=& -\overline{l_{aR}}\left[K^{\eta T }_{LRab}\eta^0
    +\frac{1}{2\Lambda} \left(K^{sT}_{LR}+K^s_{LR}\right)_{ab} \chi^{0*}\rho^{0*}\right]l_{bL}+H.c. \nonumber\\
    &\approx& -\overline{l_{aR}}\left[K^{\eta T}_{RLab}\eta^0
    +\frac{v^*_\chi}{\Lambda} \big(K^{sT}_{LR}+K^s_{LR}\big)_{ab} \rho^{0*}\right]l_{bL}+H.c. \nonumber\\
    &=& -\overline{l}_a\mathcal{H}_{iab}l_bh_i^0+i\overline{l}_a\gamma_5\mathcal{A}_{iab}l_bA_i^0 \ ,
\label{neutros}
\end{eqnarray}
where
\begin{eqnarray}
K^{\eta T}_{LR} &\equiv& V^{lT}_LG^\eta V^{l*}_R
\simeq\left(\begin{array}{ccc}
-0.0143978 & -7.36525   & 12.7554    \\
7.36067    & -0.0158071 & -9.11514   \\
-12.7351   & 9.14364    & 0.00479483 \\
\end{array}\right)\times 10^{-6}, \nonumber \\
K^{sT}_{LR} &\equiv& V^{lT}_L G^sV^{*l}_R
\simeq\left(\begin{array}{ccc}
9.50195  & 2.32646  & -3.95932 \\
-2.30121 & -1956.57 & 2.85661  \\
4.05155  & -2.8817  & 32904.1  \\
\end{array}\right)\times 10^{-6},
\label{a1e12}
\end{eqnarray}
and we have used $x^0=\textrm{Re}\,x^0+i\textrm{Im}\,x^0$ with $\textrm{Re}\, x^0=\sum_{i=1}^5 U^{h}_{xi}h_i^0$ and $\textrm{Im}\,x^0=\sum_{i=1}^5U^{A}_{xi}A_i^0$, where
$h_i^0(A_i^0)$ are scalar (pseudo-scalar) mass eigenstates, and $U^{h,A}$ are $4\times4$ orthogonal matrices, also
\begin{eqnarray}\label{matrix-Hi-Ai} \nonumber\\
\mathcal{H}_i  \equiv U_{\eta i}^hK_{LR}^{\eta T}+U_{\rho i}^h(K_{LR}^{s T}+K_{LR}^{s}), \quad
\mathcal{A}_i  \equiv U_{\eta i}^AK_{LR}^{\eta T}-U_{\rho i}^A(K_{LR}^{s T}+K_{LR}^{s}),
    \label{matrix-H-iAi}
\end{eqnarray}
where we have omitted flavor indices.

\section{Phenomenological consequences}
\label{sec:pheno}

The obtained matrices parametrizations allow us now to study the physical consequences of the m331 model. For that purpose we are going to compute some processes involving lepton flavor number violation $\Delta L_i=\pm 1$ with the motivation of finding hints of physics effects beyond the SM. First of all, we must say that our phenomenological results strongly depend on the values of the matrices in (\ref{vlr1}) and the related ones in Secs.~\ref{subsec:2vector} and \ref{subsec:2scalars}. In fact, we will show in Sec.~\ref{sec:3leptons} that with the numerical matrices the constraints on the masses of the extra vector bosons in the model are quite different from that other three matrices parametrizations samples given at the Appendix.

\subsection{$h^0\to l_i\bar{l}_j$ }

\begin{center}
\begin{figure}[!h]
\includegraphics[width=8cm]{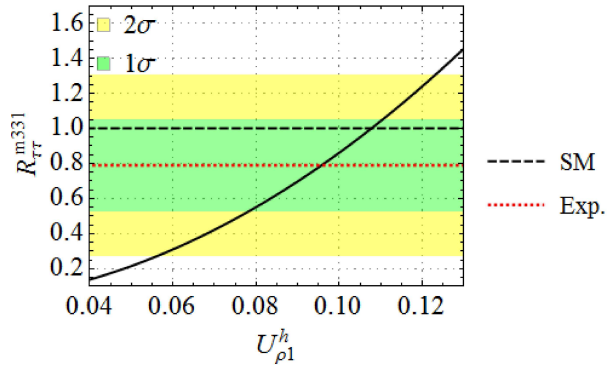}
\caption{$R_{\tau\tau}^\text{m331}$ of the decay $h^0\to\tau\bar{\tau}$, $R_{\tau\tau}^\text{m331}\equiv R_{\tau\tau}^\text{Exp}=0.79\pm 0.26$ from PDG \cite{Agashe:2014kda}. See specific values in Table \ref{TABLE-h-tautau}.}\label{FIGURE-h-tautau}
\end{figure}
\end{center}

We start our analysis by tuning our $h_1^0$ scalar with the Higgs $h^0$ of the SM, for that purpose we must
set the diagonal Yukawas in (\ref{matrix-H-iAi}) in such a way that it matches with the SM coupling given by $\frac{gm_{l}}{2m_W}$
with $l_a=e,\mu,\tau$. The well measured coupling is the $\tau$-lepton, hence for this lepton
\begin{eqnarray}\label{}
    U_{\eta 1}^hK_{LR33}^{\eta T}+U_{\rho 1}^h(K_{LR}^{s T}+K_{LR}^{s})_{33} &\approx&
    \frac{gm_\tau}{2m_W} = 9.9\times10^{-3} ,
    \label{H1-tau}
\end{eqnarray}
hence from (\ref{H1-tau})  we get
\begin{equation}
U_{\rho 1}^h \approx 0.107734 -7.28607\times10^{-8}~U_{\eta 1}^h \sim 0.107734,
\label{ur1}
\end{equation}
and, because of the suppression factor, it is independent of the value of $U_{\eta 1}^h$.
On the other hand, experimentally the $h^0\to\tau\bar{\tau}$ is the best known of the three leptonic Higgs decays \cite{Agashe:2014kda}, we use it for estimating the matrix element $U_{\rho 1}^h$ by comparing
\begin{equation}\label{}
R_{\tau\tau}^\text{m331}\equiv \frac{\Gamma(h^0\to\tau\bar{\tau})^\text{m331}}{\Gamma(h^0\to\tau\bar{\tau})^\text{SM}},
\end{equation}
with $R_{\tau\tau}^\text{Exp}=0.79\pm 0.26$ from PDG \cite{Agashe:2014kda}, here $\Gamma_{h^0}^\text{SM}=0.00407$ GeV with $m_{h^0}=125$ GeV, where we derive the central value
\begin{equation}\label{Urho1}
U_{\rho 1}^h=0.096 \ ,
\end{equation}
see Fig.~\ref{FIGURE-h-tautau} and Table~\ref{TABLE-h-tautau} for details. Replacing (\ref{Urho1}) in (\ref{matrix-H-iAi}) with $i=1$ we obtain the symmetric mixing matrix:
\begin{equation}\label{h1}
\mathcal{H}_1=
\left(
\begin{array}{ccc}
1.82437  & 0.00242367 & 0.00885405  \\
         & -375.661   & -0.00240858 \\
         &            & 6317.58     \\
\end{array}
\right)\times10^{-6} \ .
\end{equation}

Now we can predict the branching ratio of the lepton flavor violating tree level processes $h^0\to l_i\bar{l}_j=h^0\to l_i\bar{l}_j+\bar{l}_il_j$.
The width decay is
\begin{equation}\label{ij331}
\Gamma(h^0\to l_i\bar{l}_j)^\text{m331}=\frac{m_{h^0}}{8\pi}\left(|\mathcal{H}_{1ij}|^2+|\mathcal{H}_{1ji}|^2\right)
    \sqrt{\left[1-\frac{(m_{li}+m_{lj})^2}{m_{h^0}^2}\right]^3\left[1-\frac{(m_{li}-m_{lj})^2}{m_{h^0}^2}\right]} \ ,
\end{equation}
and the branching ratio
\begin{equation}\label{Br-h0-lilj}
\text{Br}(h^0\to l_i\bar{l}_j)^\text{m331} \simeq \frac{1}{\Gamma_{h^0}^\text{SM}}\frac{m_{h^0}}{4\pi}|\mathcal{H}_{1ij}|^2  .
\end{equation}
Using (\ref{h1}), from (\ref{Br-h0-lilj}) we obtain
\begin{equation}\label{tree}
\text{Br}(h^0\to\mu\bar{\tau})^\text{m331}=1.42\times 10^{-14},
\end{equation}
and we see that it is highly suppressed compared with the value in (\ref{Higgs-muontau-Exp}). The experimental data for $h^0\to\mu\bar{\tau}$ from CMS \cite{Khachatryan:2015kon} and ATLAS \cite{Aad:2015gha} have reported a very large signal, whose average value is \cite{Chakraborty:2016gff}:
\begin{equation}\label{Higgs-muontau-Exp}
\text{Br}(h^0\to\mu\bar{\tau})^\text{Exp}=8.20_{-0.32}^{+0.33}\times10^{-3}.
\end{equation}
Our complete estimations for
$h^0\to e\bar{\mu},\ e\bar{\tau},\ \mu\bar{\tau}$ are given in the Table~\ref{TABLE-h-lilj}.

Worth to mention that the $h^0\to\mu\bar{\tau}$ process can also be generated beyond tree level topologies, i. e. virtual exchange of new scalars at one-loop level, among others, which could larger than the considered tree level contributions. We will consider this processes elsewhere.

\subsection{$l_i\to l_jl_k\bar{l}_k$}
\label{sec:3leptons}

Among the possible tree level decays $l_i\to l_jl_k\bar{l}_k$, we will see that in the m331 model the channel $\mu\to ee\bar{e}$ provides a crucial test on the lepton number violating phenomenology. These processes have been studied previously in the 3-3-1 models \cite{Liu:1993gy,CortesMaldonado:2011uh,Cabarcas:2013jba}, or in extended models via vector FCNC interactions \cite{Aranda:2012qs}, but in those papers the authors do not give solution to the mixing matrices, instead they estimate bounds on the ratio of entries of the mixing matrix and the mass of the extra particles of the model.

\begin{center}
\begin{figure}[!h]
\includegraphics[width=5cm]{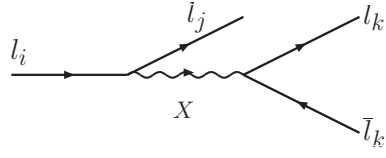}
\caption{Decay $l_i\to l_jl_k\bar{l}_k$ due to the virtual particles $X\equiv U_\mu^{++},Y^{++}, h^0,$ and $A^0$.}\label{FIGURE-3leptons}
\end{figure}
\end{center}

Although the model has several neutral scalars and pseudoscalars, here we will chose just one in each neutral sector, this means that we are considering that the other scalars are very heavy or suppressed by the matrix elements which relate the symmetry and the mass eigenstates in these sectors.
Hence, here we will consider that $l_i\to l_jl_k\bar{l}_k$ occurs only via the virtual interaction of the doubly charged vector $U_\mu^{++}$, the doubly charged scalar $Y^{++}$, the neutral scalar $h^0$, and from the neutral pseudoscalar $A_1^0$, denoted simply by $A^0$, see in Fig.~\ref{FIGURE-3leptons} the generic diagram.

The decay has the configuration $l_i(p_4)\to l_j(p_1)l_k(p_2)l_k(p_3)$, and the kinematics $p_1^2=m_{lj}^2$, $p_2^2=p_3^2=m_{lk}^2$,
$p_1\cdot p_2=m_{li}^2(-1+x+y-m_{lj}^2/m_{li}^2)/2$,
$p_1\cdot p_3=m_{li}^2(1-y-m_{lj}^2/m_{li}^2)/2$,
$p_2\cdot p_3=m_{li}^2(1-x+m_{lj}^2/m_{li}^2-2m_{lk}^2/m_{li}^2)/2$ \cite{Barger:1987nn}.

The total amplitude is conformed by four sub-amplitudes
\begin{equation}\label{Amplitude-3leptons}
\mathcal{M}=\mathcal{M}_{U^{++}}+\mathcal{M}_{Y^{++}}+\mathcal{M}_{h^0}+\mathcal{M}_{A^0} \ .
\end{equation}
In the Appendix~\ref{sec:Amplitude-3leptons} we present the example of the bilepton gauge vector contribution.
The decay width is
\begin{equation}\label{}
\Gamma(l_i\to l_jl_k\bar{l}_k)^\text{m331}=\frac{m_{li}}{256 \pi^3}
    \int_{x_\text{ini}}^{x_\text{fin}}\int_{y_\text{ini}}^{y_{\text{fin}}} |\mathcal{\overline{M}}|^2 dy dx \ ,
\end{equation}
with the mean square amplitude for the unpolarized decaying lepton $|\mathcal{\overline{M}}|^2=\frac{1}{2}\sum_\text{spin}|\mathcal{M}|^2$,
here $x$ and $y$ are dimensionless scaling variables $x\equiv 2E_j/m_{li}$, $y\equiv 2E_k/m_{li}$ \cite{Barger:1987nn}.
For this process mediated by heavy virtual particles the final lepton masses can be safely neglected without affecting the numerical results, we have verified the case with massive final leptons and the results are equivalent,
thus $x_\text{ini}=0$, $x_\text{fin}=1$, $y_\text{ini}=1-x$ and $y_\text{fin}=1$.
The resulting decay width is
\begin{eqnarray}\label{}
\Gamma(l_i\to l_jl_k\bar{l}_k)^\text{m331} &=& \Gamma_{U^{++}}+\Gamma_{Y^{++}}+\Gamma_{h^0}+\Gamma_{A^0}+\Gamma_{U^{++}-Y^{++}}+\Gamma_{U^{++}-h^0}
    +\Gamma_{U^{++}-A^0} \nonumber\\
    &&+ \Gamma_{Y^{++}-h^0}+\Gamma_{Y^{++}-A^0}+\Gamma_{h^0-A^0} \ ,
\end{eqnarray}
where the partial widths $\Gamma_X$ in terms of the leading contributions are
\begin{eqnarray}\label{partial-width-decays}
&& \Gamma_{U^{++}} = \frac{g^4 m_{li}^5}{3\cdot 2^{11}\pi^3}
    \frac{(|V_{Uij}|^2+|V_{Uji}|^2)|V_{Ukk}|^2}{m_{U^{++}}^4},  \nonumber \\&&
\Gamma_{Y^{++}} =  \frac{m_{li}^5}{3\cdot 2^7\pi^3} \left(\frac{v_\chi O_{\rho_1}}{\Lambda }\right)^4
    \frac{|K_{Rij}+K_{Rji}|^2|K_{Rkk}|^2}{m_{Y^{++}}^4} , \nonumber \\ &&
\Gamma_{h^0} = \frac{m_{li}^5}{3\cdot 2^9\pi^3}
    \frac{|\mathcal{H}_{1ij}+\mathcal{H}_{1ji}|^2|\mathcal{H}_{1kk}|^2}{m_{h^0}^4} , \nonumber \\ &&
\Gamma_{A^0} = \frac{m_{li}^5}{3\cdot 2^9\pi^3}
    \frac{|\mathcal{A}_{1ij}+\mathcal{A}_{1ji}|^2|\mathcal{A}_{1kk}|^2}{m_{A^0}^4} , \nonumber \\ &&
\Gamma_{U^{++}-Y^{++}} = 0 , \quad
\Gamma_{U^{++}-h^0} = 0 ,  \quad
\Gamma_{U^{++}-A^0} = 0 , \nonumber \\ &&
\Gamma_{Y^{++}-h^0} = \frac{m_{li}^5}{3\cdot 2^9\pi^3} \left(\frac{v_\chi O_{\rho_1}}{\Lambda }\right)^2
    \frac{(K_{Rij}+K_{Rji})K_{Rkk}(\mathcal{H}_{1ij}+\mathcal{H}_{1ji})\mathcal{H}_{1kk}}{m_{Y^{++}}^2m_{h^0}^2},\nonumber  \\ &&
\Gamma_{Y^{++}-A^0} = \frac{m_{li}^5}{3\cdot 2^9\pi^3} \left(\frac{v_\chi O_{\rho_1}}{\Lambda }\right)^2
    \frac{(K_{Rij}+K_{Rji})K_{Rkk}(\mathcal{A}_{1ij}+\mathcal{A}_{1ji})\mathcal{A}_{1kk}}{m_{Y^{++}}^2m_{A^0}^2} ,\nonumber  \\ &&
\Gamma_{h^0-A^0} = 0,
\end{eqnarray}
where $V_{U}$ is given in Eq.~(\ref{vu1}), $K_{R}$ in Eq.~(\ref{klkr}), and $\mathcal{H}_i$ and $\mathcal{A}_i$ in (\ref{matrix-Hi-Ai}).

The branching ratio is
\begin{equation}\label{branching-ratio}
\text{Br}(l_i\to l_jl_k\bar{l}_k)^{\text{m331}}=\frac{\Gamma(l_i\to l_jl_k\bar{l}_k)^\text{m331}}{\Gamma_{l_i}^\text{SM}} \ ,
\end{equation}
where the total width of the decaying lepton is obtained from its timelife $\Gamma_{l_i}^\text{SM}=1/\tau_{l_i}$.
For the $\mu$, case $\tau_\mu=2.1969811\times10^{-6}\text{s}=3.34\times10^{18}\text{GeV}^{-1}$, then $\Gamma_\mu^\text{SM}=2.99\times10^{-19}$ GeV, where 1s=1.52$\times10^{24}\text{GeV}^{-1}$;
and for the $\tau$ case $\tau_\tau=2.903\times10^{-13}\text{s}=4.41\times10^{11}\text{GeV}^{-1}$, then $\Gamma_\tau^\text{SM}=2.27\times10^{-12}$GeV. In the numerical analysis the input values are
$m_e=0.000511$ GeV, $m_\mu=0.105658$ GeV, $m_\tau=1.77682$ GeV,
$0.01\leq v_\chi O_{\rho_1}/\Lambda\leq 1$, $U_{\eta1}^h=0$, $U_{\rho 1}^h=0.096$, $U_{\eta 1}^A=0.2$, and $U_{\rho 1}^A=0.2$ already estimated in \cite{Machado:2013jca}.
For the new heavy particles masses we are going to follow the experimental bounds given in PDG~\cite{Agashe:2014kda}, for the doubly charged scalar we are going to use $m_{Y^{++}}\geq 322$ GeV, and for the pseudoescalar $m_{A^0}\geq 100$ GeV.

\begin{center}
\begin{figure}[!h]
\includegraphics[width=8.5cm]{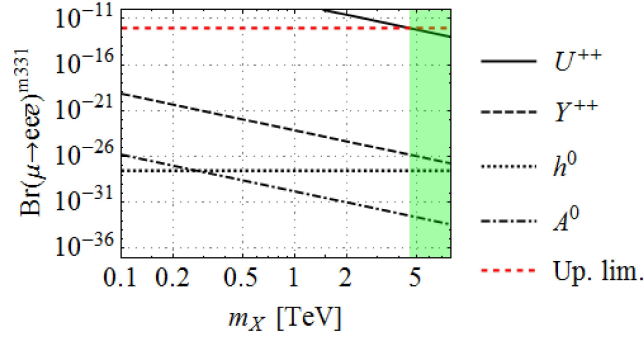}
\caption{Br$(\mu\to ee\bar{e})^\text{m331}$ showing only the partial contributions of the $U_\mu^{++}$, $Y^{++}$, $h^0$ and $A^0$,
varying simultaneously the three masses of $U_\mu^{++}$, $Y^{++}$ and $A^0$ in the same interval and for that we set them as $m_X$. The decay imposes that the mass of the $U_\mu^{++}$ vector boson must be $m_{U^{++}}>4.584$ TeV to fulfil the experimental upper limit, the green area indicates the allowed region for $m_{U^{++}}$.}\label{FIGURE-mu-eee}
\end{figure}
\end{center}

The experimental upper limit for the channel $\mu\to ee\bar{e}$ is Br$(\mu\to ee\bar{e})^\text{Exp}<10^{-12}$, see Table~\ref{Table-three-leptons}.
The Fig.~\ref{FIGURE-mu-eee} shows the decay $\mu\to ee\bar{e}$ where we are showing only the partial contributions of the $U_\mu^{++}$, $Y^{++}$, $h^0$ and $A^0$ to the Br, we vary simultaneously the three masses of $U_\mu^{++}$, $Y^{++}$ and $A^0$ in the same interval and for that we set them as $m_X$. We see from that figure the bilepton $U_\mu^{++}$ dominates entirely the process while the rest of the virtual particles are suppressed, this means that $\Gamma(\mu\to ee\bar{e})^\text{m331}\approx\Gamma_{U^{++}}$, but it must fulfill the experimental upper limit; to show this explicitly: using the elements $V_{U\mu e}$, $V_{Ue\mu}$ and $V_{Uee}$ from (\ref{vu1}) in the first line of (\ref{partial-width-decays}), we get
\begin{equation}\label{}
\text{Br}(\mu\to ee\bar{e})^\text{m331}\approx\frac{\Gamma_{U^{++}}}{\Gamma_\mu^\text{SM}}
=\frac{441.61}{m_{U^{++}}^4}\text{GeV}^4
    <\text{Br}(\mu\to ee\bar{e})^\text{Exp}=10^{-12} ,
\end{equation}
which demands $m_{U^{++}}> 4.584$ TeV. Worth to mention that the lower the values of $V_{U\mu e}$ and $V_{Ue\mu}$, the lighter $m_{U^{++}}$ could be.
Any other decay $l_i\to l_jl_k\bar{l}_k$ apart from the $\mu\to ee\bar{e}$ does not impose restrictions on any of the virtual particle masses, all those channels respect the experimental upper limits. In all the evaluations we have used
$v_\chi O_{\rho_1}/\Lambda=1$, the largest possible value for this parameter related to $Y^{++}$.
Therefore, taking advantage that in all the processes $l_i\to l_jl_k\bar{l}_k$ the $U_\mu^{++}$ bilepton absolutely domains the signal, we can neglect all the other virtual particle contributions and just focus on the $m_{U^{++}}$ dependence given in the first expression of the Eq.~(\ref{partial-width-decays}), which allow us to realise that after the numerical evaluation of the diverse matrix elements it results that for the tau decays Br$(\tau\to ee\bar{e})_\text{m331}\simeq$Br$(\tau\to e\mu\bar{\mu})_\text{m331}$ and Br$(\tau\to \mu e\bar{e})_\text{m331}\simeq$Br$(\tau\to \mu\mu\bar{\mu})_\text{m331}$, this can be appreciated in the Table~\ref{TABLE-tau-3leptons}.
In the decay $\mu\to ee\bar{e}$, besides the matrix in Eq.~(\ref{vu1}), we have also tested the three  parametrizations of the matrix $V_U$ given in Appendix \ref{sec:matrices}. See also the discussion in the Conclusions.

\subsection{$l_i\to l_j\gamma$}

In the m331 model this process is one-loop induced by the known  $W_\mu~\&~\nu_l$ with massive neutrinos \cite{W-loop,Cheng-Lee-book} and by new heavy virtual particles interacting with both virtual neutrinos and leptons. We expect that the signals coming from the virtual interaction of new particles with leptons could be larger by several orders of magnitude than the pure SM estimation, because the leptonic GIM suppression factor is $m_l^2/m_X^2 \gg m_{\nu_l}^2/m_W^2$, where $X$ denotes a new heavy bosonic particle which $m_X>m_W$.

\begin{center}
\begin{figure}[!h]
\includegraphics[width=5cm]{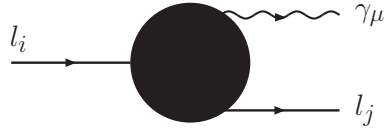}
\caption{Decay $l_i\to l_j\gamma$ due to the couple of virtual particles $W_\mu\& \nu_k,\ U_\mu^{++} \& l_k,\ V_\mu^+ \& \nu_k,\ Y^{++}\& l_k,\ h^0 \& l_k$, and $A^0 \& l_k$.}\label{FIGURE-loop}
\end{figure}
\end{center}

The decay $l_i(p_i)\to l_j(p_j)\gamma(q)$ with on-shell final states is a magnetic transition represented by a dimension five operator \cite{Cheng-Lee-book}, depicted in the Fig.~\ref{FIGURE-loop}, it has the amplitude
\begin{equation}\label{}
\mathcal{M}=\mathcal{M}^\mu\epsilon_\mu^*(\vec{q},\lambda) \ ,
\end{equation}
with kinematics $p_i=q+p_j$, $p_i^2=m_{li}^2$, $q^2=0$, $p_j^2=m_{lj}^2$, $q\cdot p_j=(m_{li}^2-m_{lj}^2)/2$,
$p_i\cdot p_j=(m_{li}^2+m_{lj}^2)/2$, and the photon transversality condition
$q^\mu\epsilon_\mu^*(\vec{q},\lambda)=0$. The Lorentz structure is
\begin{eqnarray}\label{tensor-amplitude}
\mathcal{M}^\mu &=& \bar{u}_j(p_j)i\sigma^{\mu\nu}q_{\nu}(F_M+F_E\gamma^5)u_i(p_i) \nonumber \\
    &=& \bar{u}_j(p_j)\left\{2 p_i^\mu (F_M+F_E\gamma^5)-\gamma^\mu[(m_{li}+m_{lj})F_M-(m_{li}-m_{lj})F_E\gamma^5] \right\}u_i(p_i) \ ,
\end{eqnarray}
where $\sigma^{\mu\nu}\equiv\frac{i}{2}[\gamma^\mu,\gamma^\nu]$, $F_M$ is the transition magnetic dipole moment and $F_E$ is the transition electric dipole moment. The tensor amplitude satisfies the Ward identity $q_\mu\mathcal{M}^\mu=0$ \cite{Cheng-Lee-book}.
The decay width is
\begin{equation}\label{}
\Gamma(l_i\to l_j\gamma) = \frac{1}{16\pi m_{li}}\left(1-\frac{m_{lj}^2}{m_{li}^2}\right)|\mathcal{\overline{M}}|^2
    = \frac{m_{li}^3}{8\pi}\left(1-\frac{m_{lj}^2}{m_{li}^2}\right)^3\left(|F_M|^2+|F_E|^2\right) \ ,
\end{equation}
with the mean squared amplitude
\begin{equation}\label{}\label{}
|\mathcal{\overline{M}}|^2 = \frac{1}{2}\sum_\text{spin}|\mathcal{M}|^2
    = 2m_{l i}^4\left(1-\frac{m_{l j}^2}{m_{l i}^2}\right)^2\left(|F_M|^2+|F_E|^2\right) \ .
\end{equation}
In the m331 the branching ratio is given by
\begin{equation}\label{}
\text{Br}(l_i\to l_j\gamma)^\text{m331}=\frac{\Gamma(l_i\to l_j\gamma)^\text{m331}}{\Gamma_{l_i}^\text{SM}}.
\end{equation}

Specifically, in the m331 model this process is induced by eight virtual contributions, where $W_\mu$, $V_\mu^+$ and $Y_{1,2}^+$ interact with neutrinos, and where $U_\mu^{++}$, $Y^{++}$, $h^0$ and $A^0$ interact with leptons. Nevertheless, as already mentioned, the leptonic GIM suppression factors are $m_{l}^2/m_X^2 \gg m_{\nu_{l}}^2/m_W^2 \gg m_{\nu_{l}}^2/m_X^2$, with $X$ denoting a new heavy particle of the m331 model which $m_X>m_W$, in other words, any contribution due to $X \&\ \nu_l$ is more suppressed than $W\& \ \nu_l$, for that reason we are going to omit the new cases involving neutrinos. Then, the resulting amplitude is conformed by five sub-amplitudes
\begin{equation}\label{Amplitude-loops}
\mathcal{M}=\mathcal{M}_W+\mathcal{M}_{U^{++}}+\mathcal{M}_{Y^{++}}+\mathcal{M}_{h^0}+\mathcal{M}_{A^0} \ .
\end{equation}
In the Appendix \ref{sec:Amplitude-loops} we present the sample of the $U_\mu^{++} \& \ l$ contribution.
In the loop integrals we neglect the final lepton mass, thus
\begin{eqnarray}\label{}
\Gamma(l_i\to l_j\gamma) &=& \frac{m_{li}^3}{8\pi}\left(|F_M|^2+|F_E|^2\right) \ ,
\end{eqnarray}
with the transition dipole moments
\begin{eqnarray}
F_{M,E} &=& \sum_X F_{M,E}^X \ ,
\end{eqnarray}
here $X=W_\mu$, $U_\mu^{++}$, $Y^{++}$, $h^0$, $A^0$, they  are expressed in terms of the GIM suppression factors $m_k^2/m_X^2$, with $k$ denoting in general the virtual neutrino $\nu_{lk}$ for the $W_\mu$ case or the virtual lepton $l_k$ which interacts with the new heavy particles, this fraction of masses is also known as the Inami-Lim terms. Keeping only the linear mass term $m_{li}$ \cite{Cheng-Lee-book} they are:
the known $W_\mu~\&~\nu_{lk}$
\begin{equation}\label{SM-loop}
F_{M,E}^W=-\frac{ieg^2m_{li}}{256\pi^2m_W^2}\sum_{k=1}^3V_{ik}^*V_{jk}
    \frac{m_{\nu_{lk}}^2}{m_W^2} \ ,
\end{equation}
where the matrix $V_U$ is that in (\ref{vu1}); the doubly charged vector $U_\mu^{++}~\&~l_k$
\begin{equation}\label{U2-loop}
F_{M,E}^{U^{++}} = -\frac{ieg^2m_{li}}{64\pi^2m_{U^{++}}^2}\sum_{k=1}^3(V_{Uki}V_{Ukj}^*\pm V_{Uik}V_{Ujk}^*)
    \frac{m_{lk}^2}{m_{U^{++}}^2} \ ;
\end{equation}
the doubly charged scalar $Y^{++}~\&~l_k$
\begin{equation}
F_{M,E}^{Y^{++}}=\mp\frac{iem_{li}}{16\pi^2m_{Y^{++}}^2} \left(\frac{v_\chi O_{\rho 1}}{\Lambda}\right)^2
    \sum_{k=1}^3\left(K_{Rki}^*K_{Rkj}+K_{Rik}^*K_{Rjk}\right)
    \frac{m_{lk}^2}{m_{Y^{++}}^2}\left[1+\log\left(\frac{m_{lk}^2}{m_{Y^{++}}^2}\right) \right] \ ,
\end{equation}
where $K_R$ is given in (\ref{klkr}); the neutral scalar $h^0~\&~l_k$
\begin{eqnarray}
F_M^{h^0} &=& \frac{iem_{li}}{32\pi^2m_{h^0}^2}\sum_{k=1}^3
    \left(\mathcal{H}_{ki}\mathcal{H}_{jk}+\mathcal{H}_{ik}\mathcal{H}_{kj}\right)
    \frac{m_{lk}^2}{m_{h^0}^2}\left[1+\log\left(\frac{m_{lk}^2}{m_{h^0}^2}\right) \right] \ , \nonumber\\
F_E^{h^0} &=& 0 \ ;
\end{eqnarray}
and, finally, the pseudoscalar $A^0~\&~l_k$
\begin{eqnarray}
F_M^{A^0} &=& -\frac{iem_{li}}{32\pi^2m_{A^0}^2}\sum_{k=1}^3
    \left(\mathcal{A}_{ki}\mathcal{A}_{jk}+\mathcal{A}_{ik}\mathcal{A}_{kj}\right)
    \frac{m_{lk}^2}{m_{A^0}^2}\left[1+\log\left(\frac{m_{lk}^2}{m_{A^0}^2}\right) \right] \ , \nonumber\\
F_E^{A^0} &=& 0,
\end{eqnarray}
where the matrices $\mathcal{H}$ and $\mathcal{A}$ are given in (\ref{matrix-H-iAi}), and the matrix in the $W_\mu~\&~\nu_{lk}$ case is the PMNS.

For the numerical analysis we first take into account the result of $m_{U^{++}}>4.584$ TeV derived from $\mu\to ee\bar{e}$, therefore in the following we will use $m_{U^{++}}=4590$ GeV. In the previous analyzed processes $l_i\to l_jl_k\bar{l}_k$ all of them were absolutely dominated by the $U_\mu^{++}$ bilepton mass, and therefore there was not necessary to consider the other heavy virtual particle masses, but for $l_i\to l_j\gamma$ that is not the case, in fact here the $U_\mu^{++}$ bilepton has more suppressed contribution than other particles. As in the analysis of $l_i\to l_jl_k\bar{l}_k$, we use $m_{Y^{++}}=322$ GeV, $m_{A^0}\geq 100$ GeV and $m_{h^0}=125$ GeV. The other important variable present in the decay comes from the $\overline{l_{a}^c}l_{b}Y^{++}$ interaction, which we are going to explore in the range $0.01\leq v_\chi O_{\rho_1}/\Lambda \leq 1$.

\begin{center}
\begin{figure}[!h]
\subfloat[]{\includegraphics[width=8cm]{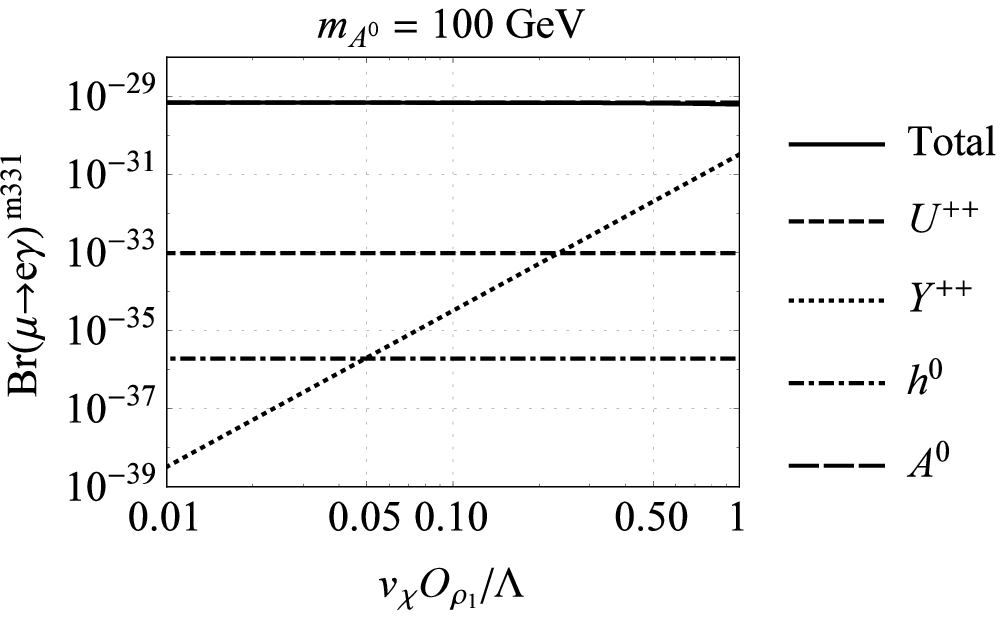}} \quad
\subfloat[]{\includegraphics[width=8cm]{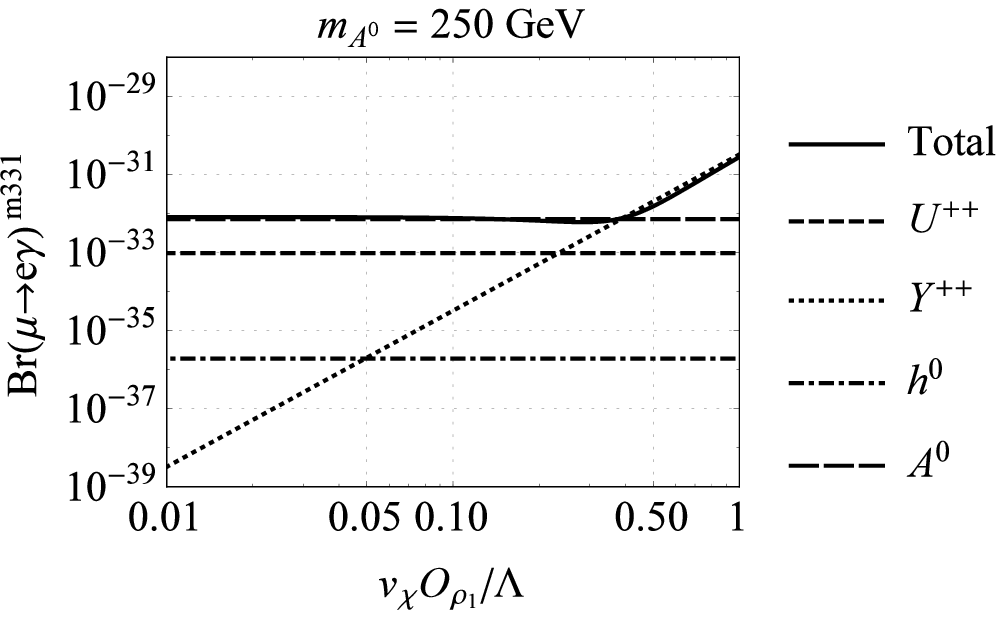}} \\
\subfloat[]{\includegraphics[width=8cm]{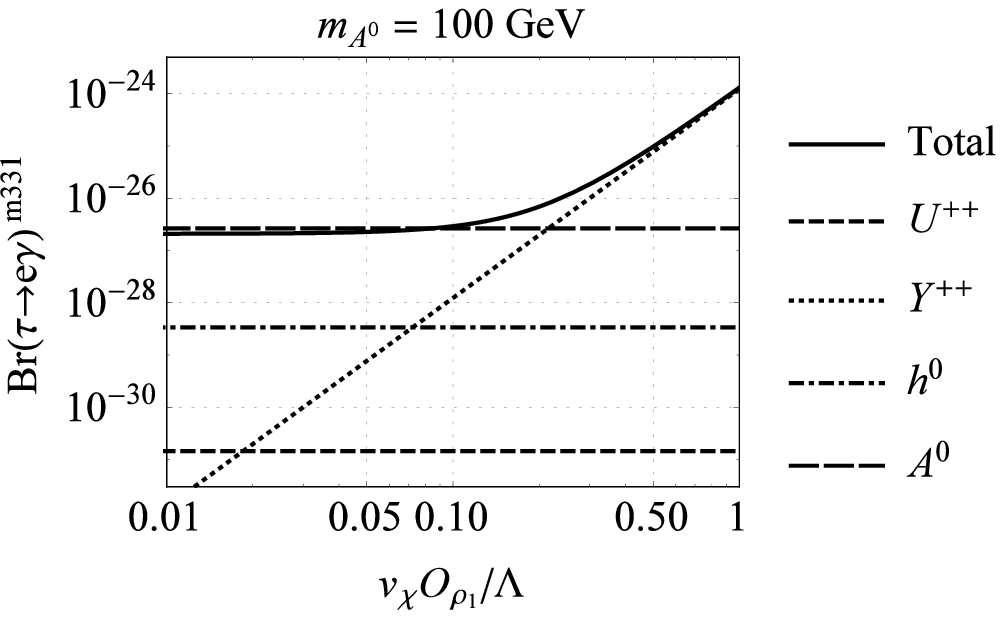}} \quad
\subfloat[]{\includegraphics[width=8cm]{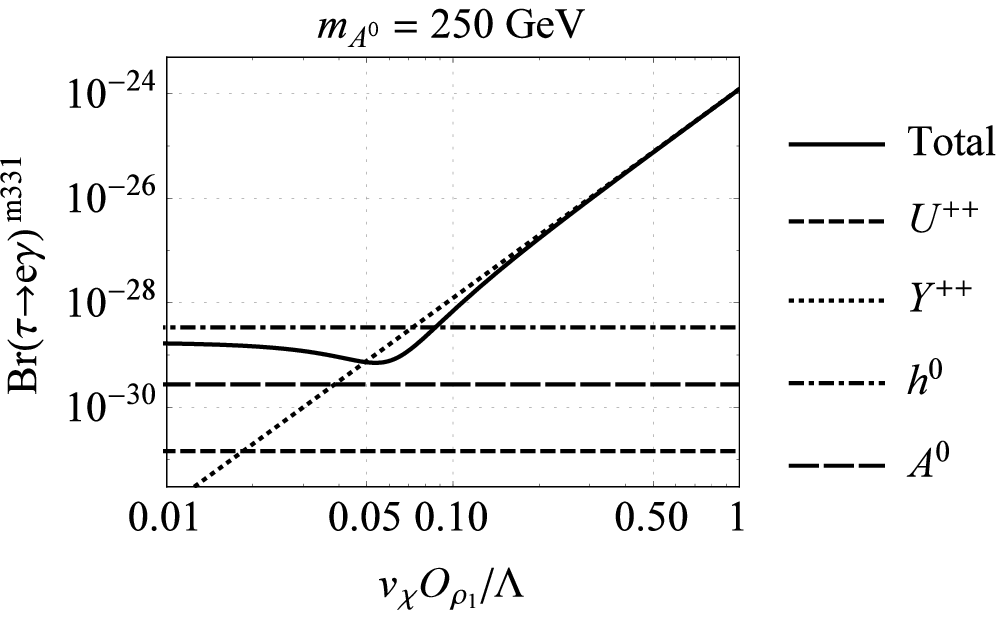}} \\
\subfloat[]{\includegraphics[width=8cm]{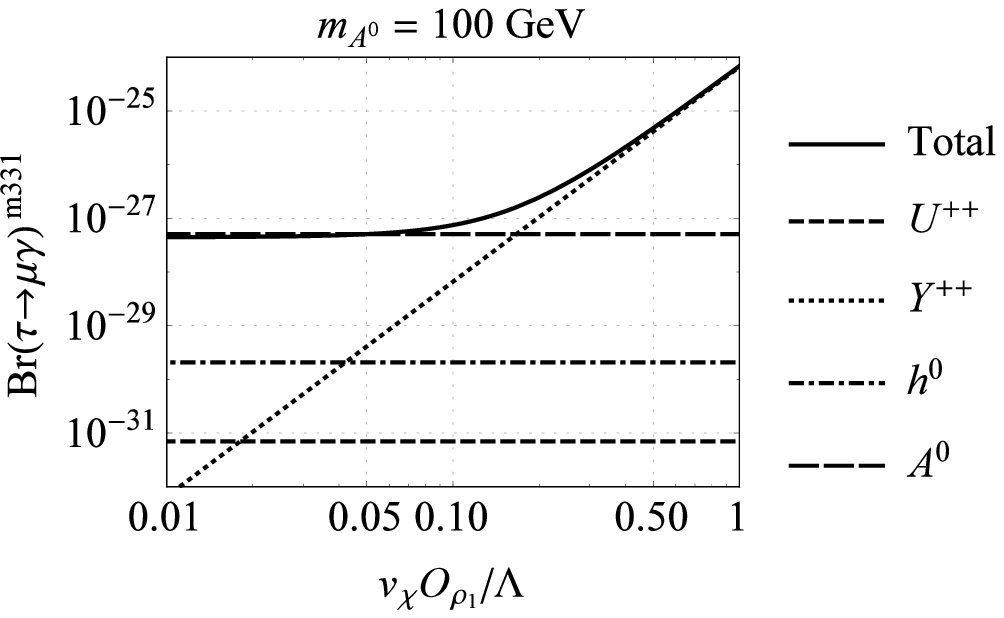}} \quad
\subfloat[]{\includegraphics[width=8cm]{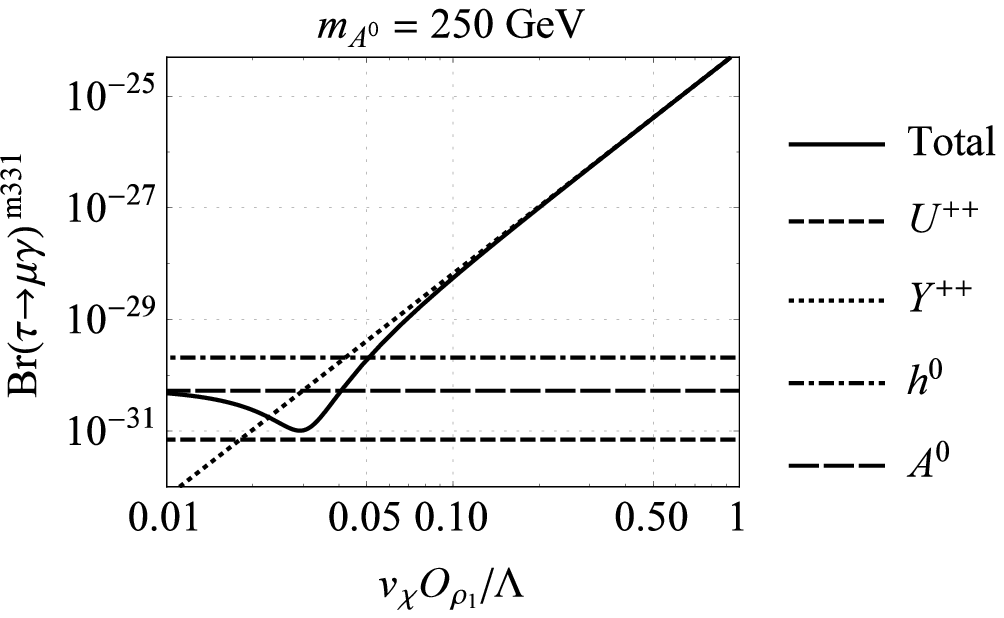}}
\caption{Br$(l_i\to l_j\gamma)^\text{m331}$ with $m_{U^{++}}=4590$ GeV, $m_{Y^{++}}=322$ GeV, with $m_{A^0}$ fixed cases and as a function of $0.01\leq v_\chi O_{\rho_1}/\Lambda\leq 1$.}\label{FIGURE-loops-Sol2}
\end{figure}
\end{center}

The Fig.~\ref{FIGURE-loops-Sol2} shows the behaviour of the three decays $\mu\to e\gamma$, $\tau\to e\gamma$ and $\tau\to\mu\gamma$ in this model, the signals of all of them are quite suppressed respect to the current experimental upper limits shown in Table~\ref{Table-loop-experiments}, but a lot of orders of magnitude greater than the SM estimation. In all these plots the total curves include also the interference among the different virtual particle contributions, but we do not plot explicitly the interference in order to not overwhelm with many curves. Specifically, the channel $\mu\to e\gamma$ is presented in the Fig.~\ref{FIGURE-loops-Sol2}(a) with $m_{A^0}=100$ GeV as function of $0.01\leq v_\chi O_{\rho_1}/\Lambda \leq 1$, where Br$^\text{m331}\sim10^{-30}$ is due entirely to the pseudoscalar $A^0$ (that is why the total contribution is the same and overlaps the $A^0$ signal), the participation of the rest of the particles are suppressed, and in the Table~\ref{TABLE-muon-egamma} can be appreciated some values in detail for given scenarios, being noticeable that our prediction is quite far from the SM estimation of Br$^\text{SM}\sim 10^{-48}$; the Fig.~\ref{FIGURE-loops-Sol2}(b) shows $m_{A^0}=250$ GeV, here the signal diminishes to Br$^\text{m331}\sim10^{-33}$ when $v_\chi O_{\rho_1}/\Lambda \lesssim 0.3$ due to the pseudosalar $A^0$, and up to Br$^\text{m331}\sim10^{-31}$ when $v_\chi O_{\rho_1}/\Lambda=1$ but the signal is now holded by the scalar $Y^{++}$. For $\tau\to e\gamma$ the Fig.~\ref{FIGURE-loops-Sol2}(c) shows that when $m_{A^0}=100$ GeV its branching ratio can goes from $10^{-24}$ to $10^{-27}$, but here the signal in the region $v_\chi O_{\rho_1}/\Lambda \lesssim 0.1$ is constant due to the pseudoscalar $A^0$, and when $0.1>v_\chi O_{\rho_1}/\Lambda=1$ the signal grows dominated by $Y^{++}$, see Table~\ref{TABLE-tau-egamma} for specific values; the case $m_{A^0}=250$ GeV is presented in the Fig.~\ref{FIGURE-loops-Sol2}(d) where Br$^\text{m331}\sim10^{-29}$ due to $h^0$ when $v_\chi O_{\rho_1}/\Lambda \lesssim 0.07$, but can reach up to Br$^\text{m331}\sim10^{-24}$ because of $Y^{++}$ if $v_\chi O_{\rho_1}/\Lambda =1$. Finally, for $\tau\to\mu\gamma$ in the Fig.~\ref{FIGURE-loops-Sol2}(e) we can see that the signal varies from Br$^\text{m331}\sim10^{-28}$ due to $A^0$ when $v_\chi O_{\rho_1}/\Lambda<0.05$, up to Br$^\text{m331}\sim10^{-25}$ owing to $Y^{++}$ if $v_\chi O_{\rho_1}/\Lambda=1$; for $m_{A^0}=250$ GeV the Fig.~\ref{FIGURE-loops-Sol2}(f) shows that $h^0$ is responsible for Br$^\text{m331}\sim10^{-30}$ if $v_\chi O_{\rho_1}/\Lambda\lesssim 0.04$, and after
$v_\chi O_{\rho_1}/\Lambda \gtrsim 0.04$ the signal grows rapidly because of $Y^{++}$ being able to reach Br$^\text{m331}\sim10^{-25}$ when $v_\chi O_{\rho_1}/\Lambda=1$.

Summarizing, our predictions for $l_i\to l_j\gamma$ are several orders of magnitude larger than the respective estimations within the pure SM due to $W \&~\nu_l$, this behavior is possible thanks to the presence of the virtual charged leptons coupling with the new heavy content of the m331 model.

\section{conclusions}
\label{sec:con}

After adjusting the masses and the unitary matrices $V^l_{L,R}$ and $V^\nu_L$ in (\ref{vlr1}) and (\ref{vlnus1}), respectively, we are left with the following free parameters, $\Lambda_s$, which is related with the mass scale of the scalar sextet and the matrices relating the mass and symmetry eigenstates in the scalar sectors: $O$ appearing in (\ref{ss2}) in the doubly charged sector, $U^h$ in the CP even sector, and $U^A$ in the CP odd sector both appearing in (\ref{matrix-H-iAi}).
Next, we were able to identify the SM Higgs $h^0$ from (\ref{neutros}) and (\ref{ur1}), which from the experimental data for $h^0\to\tau\bar{\tau}$ \cite{Agashe:2014kda} allowed us to determine $U_{\rho 1}^h=0.096$, while the parameter $U_{\eta^01}^h$ is not important, see (\ref{ur1}). Hence, the tree level flavor number violating Higgs decays are Br$(h^0\to e\bar{\mu}\ ,\mu\bar{\tau})^\text{m331}\sim10^{-14}$ and Br$(h^0\to e\bar{\tau})^\text{m331}\sim10^{-13}$, being highly suppressed respect to the reported data of Br$(h^0\to\mu\bar{\tau})^\text{Exp}\sim10^{-3}$ \cite{Khachatryan:2015kon,Aad:2015gha,Chakraborty:2016gff}. This decay also could be generated via loop interactions of the SM Higgs with new possible virtual scalars.

In the flavor number violating processes $l_i\to l_jl_k\bar{l}_k$, the channel $\mu\to ee\bar{e}$ imposed the bound $m_{U^{++}}>4.584$ TeV respecting the experimental upper limit of Br$(\mu\to ee\bar{e})^\text{Exp}<10^{-12}$, hence if in future experiments this channel is observed with a branching ratio in the range $10^{-14}-10^{-12}$ our vector bilepton $U_\mu^{\pm\pm}$ could explain it. For the tau decays we estimate for all of the reactions Br$(\tau\to l_jl_k\bar{l}_k)^\text{m331}\sim10^{-15}$ using  $m_{U^{++}}=4590$ GeV, which have resulted 7 orders of magnitude suppressed respect to the experimental upper limits.

Regarding to the one-loop level processes $l_i\to l_j\gamma$, the Br$(\mu\to e\gamma)^\text{m331}\sim10^{-33}-10^{30}$, this is up to 18 orders of magnitude larger than the SM estimation but 17 orders of magnitude below the experimental upper limit; and similar behaviour for the tau decays being Br$(\tau\to e\gamma)^\text{m331}\simeq10^{-29}-10^{-24}$ and
Br$(\tau\to \mu\gamma)^\text{m331}\simeq10^{-31}-10^{-25}$, and in contrast to the channels $l_i\to l_jl_k\bar{l}_k$ where the $U_\mu^{++}$ vector bilepton was responsible for the signals, in these one-loop processes the vector bilepton provided, in most of the cases,  the more suppressed contribution of the considered new particles interacting with leptons.

In order to verify how these predictions depend on the numerical values for the entries of those unitary matrices, we have considered in the decay $\mu\to ee\bar{e}$ different parametrizations of the matrix $V_U$ given
in Appendix~\ref{sec:matrices}. With the first of them in Eq.~(\ref{parametrization-1}), we obtain the lower limit $m_{U^{++}} > 51.8$ TeV;
the second parametrization in Eq.~(\ref{parametrization-2}), also adjusts the lepton masses and the PMNS and predicts a $m_{U^{++}} > 16.49$ TeV; the third parametrization in Eq.~(\ref{parametrization-3}), predicts $m_{U^{++}} > 3.34$ TeV, although in this case we were not able to fit a respective $V_L^\nu$ that adjust a realistic PMNS. However, since the matrix $V_L^\nu$ does not participate in the decay $\mu\to ee\bar{e}$, we have included this parametrization to exemplify how lower bound on the vector bilepton mass can be obtained.
Notice that, the more diagonal $V_U$, the lighter $m_{U^{++}}$.
It worth noting that the matrix in Eq.~$(\ref{vu1})$, which implies a lower bound on the vector bilepton mass of $m_{U^{++}} > 4.584$ TeV, it is enough to be produced at LHC, although to the best of our knowledge there has not been searches for this kind of particle. Notwithstanding, searches for quarks with exotic charges has been done at CMS \cite{Chatrchyan:2013wfa}.

The decays $\mu\to ee\bar{e}$ and $\mu\to e\gamma$ can also be considered in the 3-3-1 model with right-handed neutrinos (331RN by short)
of the sort proposed in Refs.~\cite{Montero:1992jk,Dong:2008sw}, i.e., when the leptons are in triplets $\psi_L=(\nu_l ,l, \nu^c)^T_L\sim(\textbf{3},-1/3)$. In the latter model only three triplets as $\eta,\rho,\rho^\prime$ are needed to break the gauge symmetry and give correct masses to all fermions in the model. However, it was shown in Ref.~\cite{Dong:2008sw} that in this model the processes above are suppressed as in the standard model unless a sextet is added giving also a natural small masses for neutrinos. We note that in that model there is no doubly charged vector boson and the lepton flavor violating processes are mediated only  by the doubly charged Higgs scalar in the sextet.

In the present model the $(\beta\beta)_{0\nu}$ may be induced by three mechanism: i) the Majorana mass of the light active neutrinos; ii) the Majorana mass of the heavy neutrinos, and iii) by the lepton number ($L$) violating interactions in the scalar potential.
In case i) the effective mass parameter to which the amplitude of the decay is proportional is given by
\begin{equation}
m_{\beta\beta}=\vert (V_{PMNS})^2_{ek}\,m_k\vert,
\label{bb0nus}
\end{equation}
where the $V_{PMNS}$ used is the one in the Eq.~(\ref{pmns1}) and we obtain, ignoring Majorana phases, $m_{\beta\beta}=2.5-5.05$ meV for the case of normal mass hierarchy, and $34-39$ meV when the inverse mass hierarchy is used. This occurs in other 3-3-1 models~\cite{Hernandez:2015tna}. These values are compatible with the experimental upper limit 140-380 meV~\cite{Auger:2012ar}. For heavy neutrinos their effects on the decay is suppressed by the large masses $\sim 1$ TeV, and also by the small mixing angles in their interactions with charged leptons. In principle the decay can be induced by terms like $f_3\eta^-_1\eta^-_1S^{++}_1$ in the scalar potential~\cite{Mohapatra:1981pm}. This sort of interactions breaks explicitly the total lepton number $L$ by two units and induce a contribution to the $(\beta\beta)_{0\nu}$ decay. However, we are working in the context of Ref.~\cite{DeConto:2015eia} in which terms violating $L$ are forbidden by discrete symmetries.
In this case the VEV of the sextet which would induce a Majorana mass to the active neutrinos vanishes and it is stable under quantum corrections. However, even if we allow those interactions to be present in the scalar potential, their contributions to $(\beta\beta)_{0\nu}$ according to~\cite{Schechter:1981bd,Wolfenstein:1982bf} are negligible. However, the arguments in these references assume that neutrinos gain mass from the VEV of the triplet, while in our model they are light because of the type-I seesaw mechanism. Besides the m331 model is intrinsically a multi-Higgs model and the situation is also different from that when there are a doublyt and a triplet of $SU(2)$. In particular, if the vertex $f_3\eta^-_2\eta^-_2 S^{++}_2$, where $S^{++}_2$ is a singlet of $SU(2)$, is allowed, and $C\!P$ violated the mixing $\eta^-_1-\eta^+_2$ induce a contribution to  the $(\beta\beta)_{0\nu}$ decay like the doubly charged scalar singlet of Ref.~\cite{Schechter:1981bd} which is not suppressed and may be of the order of the standard diagram which is proportional to $g^4m_\nu/M^4_W\langle p^2\rangle$. The fact that when neutrinos have Dirac and Majorana masses may evade the suppressions in the one doublet and one triplet model was pointed out in Ref.~\cite{Escobar:1982ec}.
This model has also contributions to $\mu-e$ conversion \cite{Marciano:2008zz,Wu:2016gjv}, and muonium-antimuonium conversions \cite{Pleitez:1999ix,Bernstein:2013hba}. In the latter case the lower limit for the vector bilepton mass is 850 GeV \cite{Willmann:1998gd}. These issues will be consider elsewhere.

\acknowledgments

ACBM thanks  CAPES for financial support.  JM thanks to FAPESP for financial support under the processe number  2013/09173-5. VP thanks CNPq  for partial support.

\newpage

\appendix

\section{Matrices}
\label{sec:matrices}

In Sec.~\ref{subsec:2vector} we presented in Eq.~(\ref{vu1}) one parametrization of $V_U$, and as we said in the Conclusions, this allows a $m_{U^{++}}$ from $\mu\to ee\bar{e}$ that is sufficiently small to be produced at the LHC. Below we present three more parametrizations and their impact on the lower bound of $m_{U^{++}}$ from the same decay.

\subsection{First parametization}
\label{subsec:um}

It has been shown in Ref.~\cite{DeConto:2015eia} that assuming the following Yukawa couplings
 $G^s_{11}=-0.0453,G^s_{12}=-0.0076,G^s_{13}=-0.0008,G^s_{22}=0.0015,
G^s_{23}=0.0001,G^s_{33}=1.84\times10^{-5}$, and
$G^\eta_{12}=G^\eta_{13}=G^\eta_{13}=-0.00001$, it is possible to obtain the appropriate masses for charged leptons, neutrinos and the PMNS matrix. They give the following numerical values for the unitary matrices that diagonalize the mass matrices \cite{DeConto:2015eia}:
\begin{equation}
V^l_L\approx\left(\begin{array}{ccc}
-0.0099&0.0146 &-0.9998 \\
-0.3185& -0.9479 &-0.0107 \\
0.9479&-0.3183 & -0.0140\\
\end{array}\right)\ , \
V^l_R\approx\left(\begin{array}{ccc}
0.0050&0.0072 & 0.9999\\
0.0026&0.9910  & -0.0072\\
0.9999 &-0.0027 & -0.0050\\
\end{array}\right).
\label{a1e1}
\end{equation}
From $V_U\equiv V_R^{lT}V_L^l$ we get
\begin{equation}\label{parametrization-1}
V_U\approx\left(
\begin{array}{ccc}
 0.946981 & -0.320728 & -0.0190221 \\
 -0.321056 & -0.946893 & -0.0177935 \\
 -0.012305 & 0.0229573 & -0.999661 \\
\end{array}
\right) ,
\end{equation}
and from $V_V=(V_U-V_U^T)/2$ and $V_A=(V_U+V_U^T)/2$, see the third line in Eq.~(\ref{uu2}), we have
\begin{equation}\label{}
V_V\approx\left(
\begin{array}{ccc}
 0 & 0.000163733 & -0.00335856 \\
 -0.000163733 & 0 & -0.0203754 \\
 0.00335856 & 0.0203754 & 0 \\
\end{array}
\right)
\ , \
V_A\approx\left(
\begin{array}{ccc}
 0.946981 & -0.320892 & -0.0156636 \\
 -0.320892 & -0.946893 & 0.0025819 \\
 -0.0156636 & 0.0025819 & -0.999661 \\
\end{array}
\right) .
\end{equation}
Using these matrices in $\mu\to ee\bar{e}$ we obtain the lower limit $m_{U^{++}}>51.8$ TeV.

For the neutrinos Yukawa couplings are the following: $G^s_{\nu 11} = 0.0029$, $G^s_{\nu  12} = -0.0019$,
$G^s_{\nu  13} =0.0009$,\\$G^s_{\nu  22} = 0.002$, $G^s_{\nu  23} = -0.0013$, $G^s_{\nu 33} =0.0009$, and considering  $V_{PMNS}=V_L^{l\dag}V_L^\nu$ we obtain:
\begin{equation}
V_L^\nu\approx\left(\begin{array}{ccc}
0.1943& 0.6793& 0.7077\\
 0.6455& 0.4547& -0.6137\\
 0.7386& -0.5760& 0.3502\\
\end{array}\right)
\ , \
|V_{PMNS}|\approx\left(\begin{array}{ccc}
0.82&0.55& 0.17\\
 0.512& 0.56 & 0.65\\
 0.26& 0.62& 0.74\\
 \end{array}\right).
\label{a1e2}
\end{equation}

\subsection{Second parametrization}
\label{subsec:dois}

We have found another parametrization for the matrices that diagonalize the charged lepton masses with the following values for the Yukawa couplings
$G^s_{11} =5\times 10^{-8},G^s_{12} =0.000198 , G^s_{13} = 5\times 10^{-8}, G^s_{22} = 5\times 10^{-8},G^s_{23} =0.0113,G^s_{33} =0.04376$, and
$G^\eta_{12}=G^\eta_{13}=G^\eta_{13}=5\times10^{-8}$, we obtain
\begin{equation}
V^l_L\approx\left(\begin{array}{ccc}
0.983908 & 0.156151& 0.0868391\\
 0.0777852 & 0.0631974& -0.994965\\
 -0.160853 & 0.985709 & 0.0500342\\
\end{array}\right)
\ , \
V^l_R\approx\left(\begin{array}{ccc}
0.978756 & 0.186555 & 0.0850542 \\
 0.0744144 & 0.0633254 & -0.995215\\
 -0.191048 & 0.980401 & 0.0480978\\
\end{array}\right),
\label{a1e3}
\end{equation}
\begin{equation}\label{parametrization-2}
V_U\approx
\left(
\begin{array}{ccc}
 0.999525 & -0.0307812 & 0.00139563 \\
 0.0307783 & 0.999523 & 0.00224729 \\
 -0.00146417 & -0.00220327 & 0.999997 \\
\end{array}
\right) ,
\end{equation}
and
\begin{equation}\label{}
V_V=
\left(
\begin{array}{ccc}
 0 & -0.0307798 & 0.0014299 \\
 0.0307798 & 0 & 0.00222528 \\
 -0.0014299 & -0.00222528 & 0 \\
\end{array}
\right)
\ , \
V_A=
\left(
\begin{array}{ccc}
 0.999525 & -10^{-6} & -10^{-5} \\
 -10^{-6} & 0.999523 & 10^{-5} \\
 -10^{-5} & 10^{-5}  & 0.999997 \\
\end{array}
\right) .
\end{equation}
With this parametrization we obtain the lower limit $m_{U^{++}}>16.49$ TeV from $\mu\to ee\bar{e}$.

For the neutrinos Yukawa couplings are the following $(\times 10^{-9})$: $G^\nu_{11} = 0.00677$, $G^\nu_{12} = - 0.008366$, $G^\nu_{13} =-0.0070139$, $G^\nu_{22} = 0.011457$, $G^\nu_{ 23} = 0.0067482$, $G^\nu_{33} = 0.01056$,  and from  $V_{PMNS}=V_L^{l\dag}V_L^\nu$ we obtain:
\begin{equation}
V^\nu\approx\left(\begin{array}{ccc}
0.85 & 0.09& 0.52\\
0.46& 0.62 & 0.63\\
0.27 & 0.78 & 0.57\\
\end{array}\right)
\ , \
|V_{PMNS}|\approx\left(\begin{array}{ccc}
0.83 & 0.51 & 0.17\\
0.47 & 0.59 & 0.65\\
0.27 & 0.61 & 0.73\\
\end{array}\right).
\label{a1e4}
\end{equation}

\subsection{Third parametrization}
\label{subsec:terceira}

The third parametrization yields the following values for the Yukawa couplings
$G^s_{11} = 2.191\times 10^{-5},G^s_{12} = -0.0003,G^s_{13} = -0.0001,G^s_{22} = -0.03094,G^s_{23} = 0.00801,G^s_{33} = -10^{-5}$,
and $G^\eta_{12} = -10^{-6}, G^\eta_{13} = -10^{-6}, G^\eta_{23} = -0.0001$. With them we obtain $\hat{M}^{l} = \{0.000509394, 0.105448, 1.77642\}$ GeV and the diagonalization matrices are:
\begin{equation}
V^l_L\approx\left(\begin{array}{ccc}
-0.99614 & -0.08739  &-0.00826 \\
0.01357 & 0.24625 &-0.96691  \\
0.08672 & 0.96526  & 0.24649 \\
\end{array}\right) \ , \
V^l_R\approx\left(\begin{array}{ccc}
0.99624 & -0.08629  & -0.00801 \\
0.01179 & 0.226594  & -0.97392 \\
0.08586 &0.97016 & 0.22676 \\
\end{array}\right).
\label{a1e7}
\end{equation}

\begin{equation}\label{parametrization-3}
V_U\approx
\left(
\begin{array}{ccc}
 0.999997 & -0.00127907 & 0.00150715 \\
 0.00124841 & 0.999794 & 0.0202514 \\
 -0.00153271 & -0.020248 & 0.999795 \\
\end{array}
\right) ,
\end{equation}
and
\begin{equation}\label{}
V_V=
\left(
\begin{array}{ccc}
 0 & -0.00126374 & 0.00151993 \\
 0.00126374 & 0 & 0.0202497 \\
 -0.00151993 & -0.0202497 & 0 \\
\end{array}
\right)
\ , \
V_A=
\left(
\begin{array}{ccc}
 0.999997 & -10^{-5} & -10^{-5} \\
 -10^{-5} & 0.999794 & 10^{-6} \\
 -10^{-5} & 10^{-6}  & 0.999795 \\
\end{array}
\right).
\end{equation}
Using $V_U$ in Eq.~(\ref{parametrization-3}) in the decay $\mu\to ee\bar{e}$ we obtain the lower limit $m_{U^{++}}>3.34$ TeV.

\newpage

\section{Amplitudes of the decays}
\label{sec:amplitudes}

For the computing of the amplitudes involving fermion number violating interactions we have follow the
algorithm of Refs.~\cite{Denner:1992vza,Denner:1992me} which allows the great advantage of constructing amplitudes with Feynman rules without the explicit charge conjugation matrix $C$, we only need the common Dirac propagator and less vertices than in the conventional treatment \cite{Jones:1983eh,Haber:1984rc,Gates:1987ay,Gluza:1991wj}.
The algorithm is summarized as: given the lagrangian $\mathcal{L}=\bar{\chi}\Gamma\chi=g_{abc}^i\bar{\chi}_a\Gamma_i\chi_b\phi_c$, with $\chi$ a Dirac or Majorana fermion, $\Gamma$ represents a generic fermionic interaction including Dirac matrices
$\Gamma_i=\textbf{1}, i\gamma_5,\gamma_\mu\gamma_5,\gamma_\mu,\sigma_{\mu\nu}$, coupling constants $g_{abc}^i$, and $\phi_c$ denotes scalar and vector bosonic fields. Each process diagram must be constructed twice because for every fermionic vertex two Feynman rules arise: the direct one ($\Gamma$ from $\bar{\chi}\Gamma\chi$) and the reverse one ($\Gamma '=C\Gamma^TC^{-1}$ from $\bar{\chi^c}\Gamma '\chi^c$). For a pure Majorana fermion (or general charge conjugate fermion field) $\Gamma=\Gamma '$.
Since the fermion number flow is violated it is substituted by a continuous fermion flow, an (arbitray) orientation of each complete fermion chain.

\begin{center}
\begin{figure}[!h]
\subfloat[]{\includegraphics[width=4.5cm]{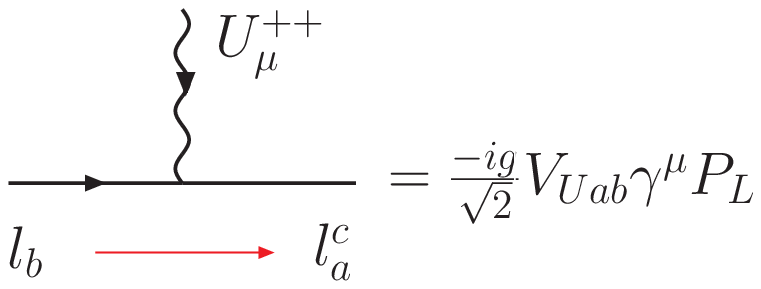}} \qquad
\subfloat[]{\includegraphics[width=4.5cm]{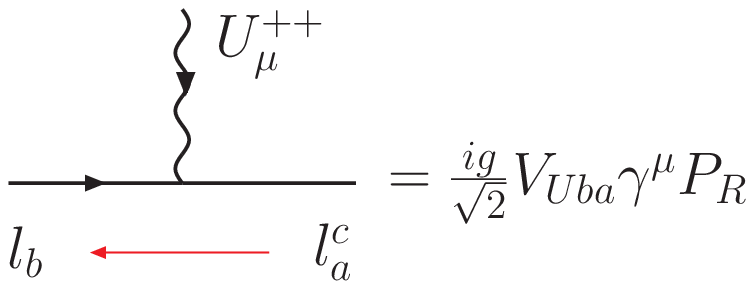}} \\
\subfloat[]{\includegraphics[width=4.5cm]{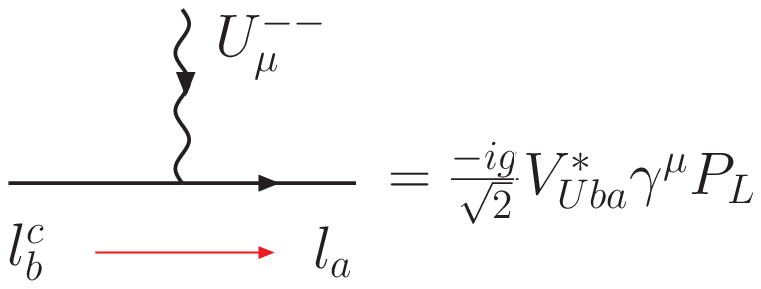}} \qquad
\subfloat[]{\includegraphics[width=4.5cm]{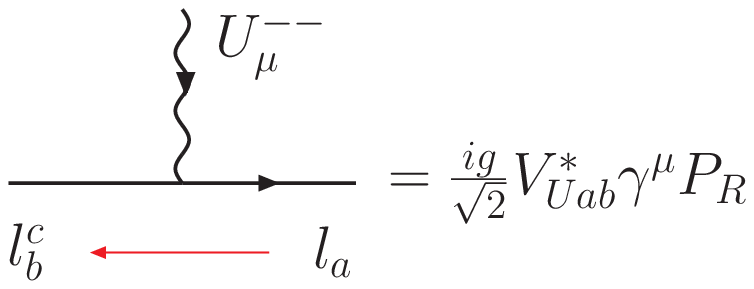}} \\
\subfloat[]{\includegraphics[width=4.5cm]{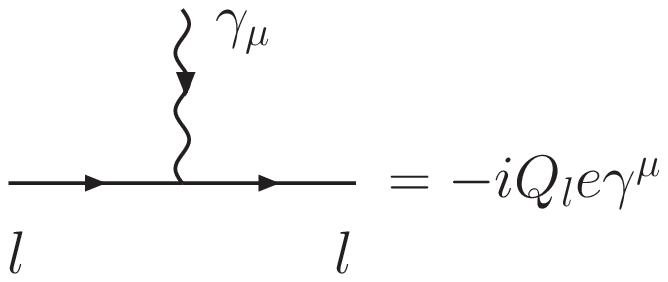}} \qquad
\subfloat[]{\includegraphics[width=4.5cm]{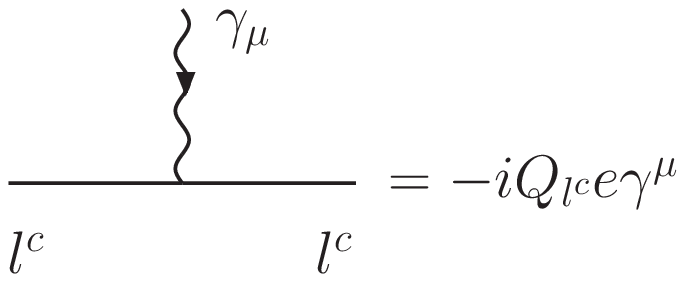}} \\
\subfloat[]{\includegraphics[width=8cm]{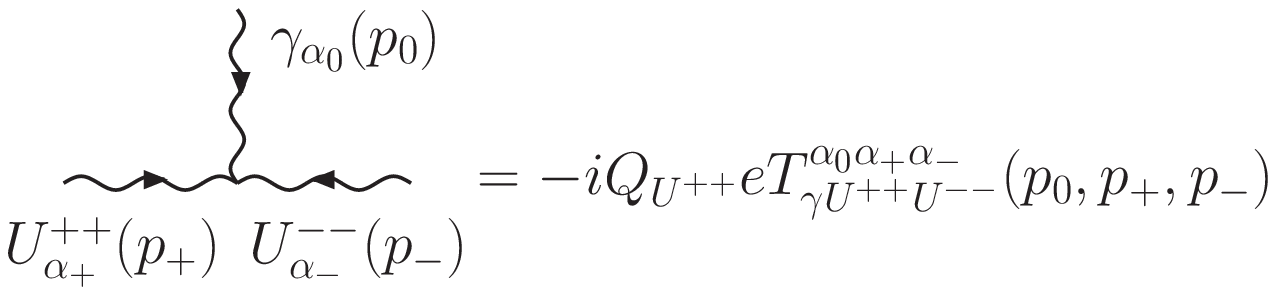}}
\caption{Feynman rules involved in the loops of the Fig.~\ref{FIGURE-3leptons-parts} and \ref{FIGURE-loops}. In the leptonic rules the left-hand side lepton is incoming and the right-hand side lepton is outgoing.}\label{FIGURE-FeynmanRules}
\end{figure}
\end{center}

Below we are going to present the samples of the vector gauge boson $U_\mu^{\pm\pm}$ contributions in each studied process.
The interactions of the $U_\mu^{\pm\pm}$ bileptons with chiral leptons were given in Eq.~(\ref{uu2}),  and although we have split the interactions in terms of left- and right-handed currents in our calculations we will use all currents as left-handed in order to use the unitary gauge, see for instance Ref.~\cite{Bu:2008fx}. Hence, from the first term in the first line of Eq.~(\ref{uu2}) and its corresponding Hermitian conjugate:
\begin{align}\label{}
\mathcal{L} =& -\frac{g}{\sqrt{2}}\left(\overline{l^c_{aL}}\gamma^\mu V_{Uab}l_{bL}U_\mu^{++}
    +\overline{l_{aL}}\gamma^\mu V_{Uab}^\dagger l_{bL}^cU_\mu^{--}\right) \nonumber\\
    =& -\frac{g}{\sqrt{2}}\left(\overline{l^c_a}\gamma^\mu P_LV_{Uab}l_bU_\mu^{++}
    +\overline{l_a}\gamma^\mu P_LV_{Uab}^\dagger l_b^cU_\mu^{--}\right) ,
\end{align}
where $V_{Uab}^\dag\equiv(V_U^\dag)_{ab}=V_{Uba}^*$, and accordingly with the algorithm \cite{Denner:1992vza,Denner:1992me}, if $\Gamma=\gamma^\mu P_L$ then $\Gamma'=C\Gamma^TC^{-1}=-\gamma^\mu P_R$, which give rise to two Feynman rules for each interaction, see Fig.~\ref{FIGURE-FeynmanRules}(a)-(d).
The photon interaction with leptons is
\begin{equation}\label{}
\mathcal{L} = -eQ_l\overline{l_a}\gamma^\mu l_aA_\mu = -e Q_{l^c}\overline{l_a^c}\gamma^\mu l_a^cA_\mu \ ,
\end{equation}
with $Q_l=-1$ and $Q_{l^c}=+1$, if $\Gamma=\gamma^\mu $ then $\Gamma'=C\Gamma^TC^{-1}=-\gamma^\mu$, whose Feynman rules are in Fig.~\ref{FIGURE-FeynmanRules}(e)-(f). The tensor definition of the vertex $\gamma U^{++}U^{--}$ in the unitary gauge, given in the Fig.~\ref{FIGURE-FeynmanRules}(g) with $Q_{U^{++}}=2$, is
\begin{equation}\label{}
T_{\gamma U^{++}U^{--}}^{\alpha_0\alpha_+\alpha_-}(p_0,p_+,p_-)\equiv(p_--p_+)^{\alpha_0}g^{\alpha_-\alpha_+}
    +(p_0-p_-)^{\alpha_+}g^{\alpha_0\alpha_-}+(p_+-p_0)^{\alpha_-}g^{\alpha_+\alpha_0} .
\end{equation}
The vector gauge boson and fermion propagators are
\begin{align}
i S_{\alpha\beta}^V(k) =& \frac{i}{k^2-m_V^2}\left(-g_{\alpha\beta}+\frac{k_\alpha k_\beta}{m_V^2} \right) \ , \\
i S_f(k) =& i\frac{\slashed{k}+m_f}{k^2-m_f^2} \ .
\end{align}

\subsection{Vector $U_\mu^{++}$ contribution to $l_i\to l_jl_k\bar{l}_k$}
\label{sec:Amplitude-3leptons}

\begin{center}
\begin{figure}[!h]
\includegraphics[width=5.5cm]{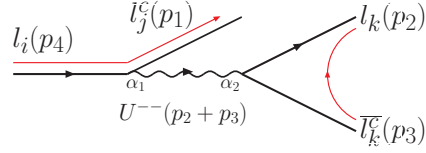}
\caption{Decay $l_i\to l_jl_k\bar{l}_k$ with the $U^{++}$ contribution.}
\label{FIGURE-3leptons-parts}
\end{figure}
\end{center}

The contribution of the $U_\mu^{++}$ bilepton to the decay $l_i\to l_jl_k\bar{l}_k$ is illustrated in  the Fig.~\ref{FIGURE-3leptons-parts}, where the red line denotes the choosen fermion flow required by the algorithm, it has the subamplitude
\begin{align}\label{}
\mathcal{M}_{U^{++}} =& \quad \left[\bar{u}_j(p_1)\left(\frac{-ig}{\sqrt{2}}V_{Uji}\gamma^{\alpha_1}P_L\right)
    u_i(p_4)\right]
    \left[iS_{\alpha_1\alpha_2}^{U^{++}}(p_2+p_3)\right] \left[\bar{u}_k(p_2)\left(\frac{-ig}{\sqrt{2}}V_{Ukk}^*\gamma^{\alpha_2}P_L\right)v_k(p_3)\right] \nonumber\\
    &+\left[\bar{u}_j(p_1)\left(\frac{ig}{\sqrt{2}}V_{Uij}\gamma^{\alpha_1}P_R\right)
    u_i(p_4)\right]
    \left[iS_{\alpha_1\alpha_2}^{U^{++}}(p_2+p_3)\right] \left[\bar{u}_k(p_2)\left(\frac{-ig}{\sqrt{2}}V_{Ukk}^*\gamma^{\alpha_2}P_L\right)v_k(p_3)\right] \nonumber\\
    =& -\frac{ig^2}{2} V_{Uji}V_{Ukk}^*
    \left[\bar{u}_j(p_1)\gamma^{\alpha_1}P_Lu_i(p_4)\right]
    \frac{1}{(p_2+p_3)^2-m_{U^{++}}^2}\left[-g_{\alpha_1\alpha_2}+\frac{(p_2+p_3)_{\alpha_1} (p_2+p_3)_{\alpha_2}}{m_{U^{++}}^2} \right] \nonumber\\
    & \times \left[\bar{u}_k(p_2)\gamma^{\alpha_2}P_Lv_k(p_3)\right] \nonumber\\
    & +\frac{ig^2}{2} V_{Uij}V_{Ukk}^* \left[\bar{u}_j(p_1)\gamma^{\alpha_1}P_Ru_i(p_4)\right]
    \frac{1}{(p_2+p_3)^2-m_{U^{++}}^2}\left[-g_{\alpha_1\alpha_2}+\frac{(p_2+p_3)_{\alpha_1} (p_2+p_3)_{\alpha_2}}{m_{U^{++}}^2} \right] \nonumber\\
    & \times\left[\bar{u}_k(p_2)\gamma^{\alpha_2}P_Lv_k(p_3)\right] \ .
\end{align}

We have solved the amplitudes with the help of \texttt{Mathematica} and \texttt{FeynCalc} \cite{Mertig:1990an,Shtabovenko:2016sxi}.

\subsection{Vector $U_\mu^{++}$ contribution to $l_i\to l_j\gamma$}
\label{sec:Amplitude-loops}

The one-loop decay $l_i\to l_j\gamma$ calculated in a renormalizable theory has finite transition magnetic and electric dipole moments, $F_M$ and $F_E$ respectively, because there are no counterterms at the tree level Lagrangian that may cancel out ultraviolet divergencies. These transition dipole form factors arise directly from triangle topologies, they can be determined from the contributions proportional to $p_i^\mu$ in Eq.~(\ref{tensor-amplitude}), therefore is sufficient to consider only these topologies to obtain the transition dipole moments. Nevertheless, to prove the electromagnetic gauge invariance and the finiteness of the process as a whole the bubbles must be considered \cite{Cheng-Lee-book,Lavoura:2003xp,Romao-book,references-with-bubbles}.
In general, the one-loop decay with on-shell final states mediated by charged bosons has triangle and bubble contributions, characterized by their respective form factors $F_T$ and $F_B$ which give rise to the amplitude
\begin{align}
\mathcal{M}^\mu=&\bar{u}_j(p_j)\left[F_{T1}p_i^\mu+F_{T2}p_i^\mu\gamma^5+(F_{T3}+F_{B3})\gamma^\mu
    +(F_{T4}+F_{B4})\gamma^\mu\gamma^5\right]u_i(p_i) \nonumber\\
    =& \bar{u}_j(p_j)\left(F_{T1}p_i^\mu+F_{T2}p_i^\mu\gamma^5-\frac{m_{li}+m_{lj}}{2}F_{T1}\gamma^\mu
    +\frac{m_{li}-m_{lj}}{2}F_{T2}\gamma^\mu\gamma^5\right)u_i(p_i) \nonumber\\
    =&\bar{u}_j(p_j)\left\{2p_i^\mu\left(\frac{F_{T1}}{2}+\frac{F_{T2}}{2}\gamma^5\right)
    +\gamma^\mu\left[-(m_{li}+m_{lj})\frac{F_{T1}}{2}
    +(m_{li}-m_{lj})\frac{F_{T2}}{2}\gamma^5\right]\right\}u_i(p_i) \nonumber\\
    =&\bar{u}_j(p_j)i\sigma^{\mu\nu}q_\nu\left(\frac{F_{T1}}{2}+\frac{F_{T2}}{2}\gamma^5\right)u_i(p_i) \ ,
\end{align}
noticing that the last two lines are precisely the Eq.~(\ref{tensor-amplitude}) with $F_{T1}/2=F_M$ and $F_{T2}/2=F_E$, besides $F_{B1,B2}=0$ and $F_{T3,B3,T4,B4}$ are divergent while $F_{T3}+F_{B3}=-(m_{li}+m_{lj})F_{T1}/2$ and $F_{T4}+F_{B4}=(m_{li}-m_{lj})F_{T2}/2$ do not.
The bubble contribution is canceled by factors coming from the triangle and all remains in terms of pure triangle information $F_{T1,2}$. When $m_{lj}=0$ occurs that $F_{T1}=F_{T2}$, then $F_M=F_E$:
\begin{equation}\label{}
\mathcal{M}^\mu=F_M\bar{u}_j(p_j)(1+\gamma^5)(2p_i^\mu-m_{li}\gamma^\mu)u_i(p_i) \ .
\end{equation}

Back to our model, in the virtual contribution $\mathcal{M}_{U^{++}}~\&~l_k$ we consider the complete set of topologies to fully prove finiteness and the Ward identity of the process, which has been crucial to us to confirm the correct application of the algorithm. The amplitude $\mathcal{M}_{U^{++}}$ of Eq.~(\ref{Amplitude-loops}) is conformed by the four diagrams depicted in the Fig.~\ref{FIGURE-loops} in the unitary gauge, the red line indicates the choosen fermion flow. We have crosschecked the vector gauge contributions using the Feynman-'t Hooft gauge and the non-linear gauge, see \cite{Montano:2005gs} and Lee-Shrock in \cite{W-loop}, proving that the transition dipole moments $F_{M,E}$ for the vector contributions are indepentent respect to the renormalization procedure, just as showed in \cite{Cheng-Lee-book} for the SM case, that accordingly with \cite{Bu:2008fx} this is true for pure left-handed couplings which is our case.
One set of diagrams, lets denote it as A, is constructed with the direct Feynman rules ($\Gamma$ from $\bar{\chi}\Gamma\chi$), and the other set B with the reverse ones ($\Gamma '=C\Gamma^TC^{-1}$ from $\bar{\chi^c}\Gamma '\chi^c$). We first compute the amplitude with $m_{lj}\neq 0$, the final massless case will be performed later. The tensor amplitude is
\begin{eqnarray}\label{ampitude-U}
\mathcal{M}_{U^{++}}^\mu &=& \mathcal{M}_{U^{++}A}^\mu+\mathcal{M}_{U^{++}B}^\mu
    = \sum_{n=1}^4\mathcal{M}_{An}^\mu+\sum_{n=1}^4\mathcal{M}_{Bn}^\mu \ ,
\end{eqnarray}
where the set A is
\begin{align}
\mathcal{M}_{A1}^\mu =& \int\frac{d^Dk}{(2\pi)^D} \bar{u}_j(p_j)
    \left(\frac{-ig}{\sqrt{2}}V_{Ukj}^*\gamma^{\alpha_1}P_L\right)
    \left[iS_{lk}(k)\right] \left(\frac{-ig}{\sqrt{2}}V_{Uki}\gamma^{\alpha_4}P_L\right)
    u_i(p_i)
    \left[iS_{\alpha_1\alpha_2}^{U^{++}}(k-p_j)\right] \nonumber\\
    & \times \left[-iQ_{U^{++}}eT^{\mu\alpha_2\alpha_3}_{\gamma U^{++}U^{--}}(-q,k-p_j,-k+q+p_j)\right]
    \left[iS_{\alpha_3\alpha_4}^{U^{++}}(k-q-p_j)\right] \nonumber\\
    =& \frac{g^2}{2}Q_{U^{++}}e \sum_{k=1}^3V_{Uki}V_{Ukj}^*\int\frac{d^Dk}{(2\pi)^D}
    \frac{\bar{u}_j(p_j)\gamma^{\alpha_1}P_L\left(\slashed{k}+m_{lk}\right)\gamma^{\alpha_4}P_Lu_i(p_i)}
    {(k^2-m_{lk}^2)[(k-p_j)^2-m_{U^{++}}^2] [(k-q-p_j)^2-m_{U^{++}}^2]} \nonumber\\
    & \times \left[-g_{\alpha_1\alpha_2}+\frac{(k-p_j)_{\alpha_1}(k-p_j)_{\alpha_2}}{m_{U^{++}}^2}\right]
    T^{\mu\alpha_2\alpha_3}_{\gamma U^{++}U^{--}}(-q,k-p_j,-k+q+p_j) \nonumber\\
    & \times\left[-g_{\alpha_3\alpha_4}+\frac{(k-q-p_j)_{\alpha_3}(k-q-p_j)_{\alpha_4}}{m_{U^{++}}^2}\right] \ ,
\end{align}
\begin{align}
\mathcal{M}_{A2}^\mu =& \int\frac{d^Dk}{(2\pi)^D}
    \bar{u}_j(p_j)\left(\frac{-ig}{\sqrt{2}}V_{Ukj}^*\gamma^{\alpha_1}P_L\right) \left[iS_{lk}(k-p_j)\right]
    (-iQ_{l^c}e\gamma^\mu)\left[iS_{lk}(k-q-p_j)\right]
    \left(\frac{-ig}{\sqrt{2}}V_{Uki}\gamma^{\alpha_2}P_L\right)u_i(p_i) \nonumber\\
    & \times\left[iS_{\alpha_1\alpha_2}^{U^{++}}(k)\right] \nonumber\\
    =& \frac{g^2}{2}Q_{l^c}e \sum_{k=1}^3V_{Uki}V_{Ukj}^* \int\frac{d^Dk}{(2\pi)^D}
    \frac{\bar{u}_j(p_j)\gamma^{\alpha_1}P_L\left(\slashed{k}-\slashed{p}_j+m_{lk}\right)\gamma^\mu
    \left(\slashed{k}-\slashed{q}-\slashed{p}_j+m_{lk}\right)\gamma^{\alpha_2}P_Lu_i(p_i)}
    {\left(k^2-m_{U^{++}}^2\right)\left[(k-p_j)^2-m_{lk}^2\right]\left[(k-q-p_j)^2-m_{lk}^2\right]} \nonumber\\
    & \times\left(-g_{\alpha_1\alpha_2}+\frac{k_{\alpha_1}k_{\alpha_2}}{m_{U^{++}}^2}\right) \ ,
\end{align}
\begin{align}
\mathcal{M}_{A3}^\mu =& \int\frac{d^Dk}{(2\pi)^D}
    \bar{u}_j(p_j)\left(\frac{-ig}{\sqrt{2}}V_{Ukj}^*\gamma^{\alpha_1}P_L\right)\left[iS_{lk}(k)\right]
    \left(\frac{-ig}{\sqrt{2}}V_{Uki}\gamma^{\alpha_2}P_L\right)\left[iS_{li}(p_j)\right]
    (-iQ_le\gamma^\mu)u_i(p_i) \left[iS_{\alpha_1\alpha_2}^{U^{++}}(k-p_j)\right] \nonumber\\
    =& \frac{g^2}{2}Q_le \sum_{k=1}^3V_{Uki}V_{Ukj}^* \frac{1}{m_{lj}^2-m_{li}^2}
    \int\frac{d^Dk}{(2\pi)^D}
    \frac{\bar{u}_j(p_j)\gamma^{\alpha_1}P_L\left(\slashed{k}+m_{lk}\right)\gamma^{\alpha_2}P_L
    \left(\slashed{p}_j+m_{li}\right)\gamma^\mu u_i(p_i)}
    {\left(k^2-m_{lk}^2\right)\left[(k-p_j)^2-m_{U^{++}}^2\right]} \nonumber\\
    & \times\left[-g_{\alpha_1\alpha_2}+\frac{(k-p_j)_{\alpha_1}(k-p_j)_{\alpha_2}}{m_{U^{++}}^2}\right] \ ,
\end{align}
\begin{align}
\mathcal{M}_{A4}^\mu =& \int\frac{d^Dk}{(2\pi)^D}
    \bar{u}_j(p_j)(-iQ_le\gamma^\mu)\left[iS_{lj}(q+p_j)
    \right]\left(\frac{-ig}{\sqrt{2}}V_{Ukj}^*\gamma^{\alpha_1}P_L\right)\left[iS_{lk}(k)\right]
    \left(\frac{-ig}{\sqrt{2}}V_{Uki}\gamma^{\alpha_2}P_L\right)u_i(p_i) \nonumber\\
    &\times\left[iS_{\alpha_1\alpha_2}^{U^{++}}(k-q-p_j)\right] \nonumber\\
    =& \frac{g^2}{2}Q_le \sum_{k=1}^3V_{Uki}V_{Ukj}^* \frac{1}{m_{li}^2-m_{lj}^2}
    \int\frac{d^Dk}{(2\pi)^D}
    \frac{\bar{u}_j(p_j)\gamma^\mu\left(\slashed{q}+\slashed{k}+m_{lj}\right)\gamma^{\alpha_1}P_L
    \left(\slashed{k}+m_{lk}\right)\gamma^{\alpha_2}P_Lu_i(p_i)}
    {\left(k^2-m_{lk}^2\right)\left[(k-q-p_j)^2-m_{U^{++}}^2\right]} \nonumber\\
    & \times\left[-g_{\alpha_1\alpha_2}+\frac{(k-q-p_j)_{\alpha_1}(k-q-p_j)_{\alpha_2}}{m_{U^{++}}^2}\right] \ ;
\end{align}
and the set B results
\begin{align}\label{}
\mathcal{M}_{B1}^\mu =& \mathcal{M}_{A1}^\mu (V_{Ukj}^*\to V_{Ujk}^*,V_{Uki}\to V_{Uik},P_L\to P_R) \ , \\
\mathcal{M}_{B2}^\mu =& \mathcal{M}_{A2}^\mu (V_{Ukj}^*\to V_{Ujk}^*,V_{Uki}\to V_{Uik},P_L\to P_R) \ , \\
\mathcal{M}_{B3}^\mu =& \mathcal{M}_{A3}^\mu (V_{Ukj}^*\to V_{Ujk}^*,V_{Uki}\to V_{Uik},P_L\to P_R) \ , \\
\mathcal{M}_{B4}^\mu =& \mathcal{M}_{A4}^\mu (V_{Ukj}^*\to V_{Ujk}^*,V_{Uki}\to V_{Uik},P_L\to P_R) \ .
\end{align}
Each set $\mathcal{M}_{U^{++}A,B}^\mu$ is finite because the ultraviolet term $\Delta\equiv 2/(4-D)-\gamma_E+\log 4\pi=1/\epsilon-\gamma_E+\log 4\pi$, $D=4-2\epsilon$, is canceled out, it arise from the Passarino-Veltman functions $B_0$ given below, and the electromagnetic gauge invariances is also satisfied $q_\mu\mathcal{M}_{U^{++}A,B}^\mu=0$.

Now we turn to consider the approximation $m_{lj}=0$ in (\ref{ampitude-U}), we get
\begin{align}\label{form-factor-U2}
F_{M,E}^{U^{++}}=& \frac{-m_{li}^2 \left(m_{lk}^2-2 m_{U^{++}}^2\right)+m_{lk}^4+m_{lk}^2 m_{U^{++}}^2
    -2 m_{U^{++}}^4}{64 m_{li}^4}[B_0^{U^{++}}(1)-B_0^{U^{++}}(2)] \nonumber\\
    & +\frac{m_{U^{++}}^2 \left(-2 m_{li}^2+m_{lk}^2+2 m_{U^{++}}^2\right)}{32 m_{li}^2}C_0^{U^{++}}(1)
    +\frac{ m_{lk}^2 \left(-m_{li}^2+m_{lk}^2+2 m_{U^{++}}^2\right)}{64 m_{li}^2}C_0^{U^{++}}(2) \nonumber\\
    & +\frac{3 \left(m_{lk}^2+2 m_{U^{++}}^2\right)}{128 m_{li}^2} \ .
\end{align}
We obtain the analytical solutions of the Passarino-Veltman scalar functions with the help of \texttt{Package-X} \cite{Patel:2015tea}, considering the approximation $m_{lk}=0$ in $B_0(1)$, $B_0(2)$ and $C_0(1)$, because $m_{lk}\ll m_X$, but in $C_0(2)$ we have set just one $m_{lk}$ to zero, later we Taylor expand each result around $m_{li}^2/m_X^2\ll 1$, and also a second expansion in $C_0(2)$ around $m_{lk}^2/m_X^2\ll 1$. Proceeding in this way we have also reproduced the known result for the SM case due to $W\&~\nu_l$ given in Eq.~(\ref{SM-loop}), see \cite{Cheng-Lee-book}. The approximations of the Passarino-Veltman functions are
\begin{align}
B_0^X(1) &\equiv B_0(0,m_{lk}^2,m_X^2)
    = -i16\pi^2\mu^{2\epsilon}\int\frac{d^Dk}{(2\pi)^D} \frac{1}{(k^2-m_{lk}^2)[(k-p_j)^2-m_X^2]} \nonumber\\
    &\approx \Delta+\log\left(\frac{\mu^2}{m_X^2}\right)+1 \ ,
\end{align}
\begin{align}
B_0^X(2) &\equiv B_0(m_{li}^2,m_{lk}^2,m_X^2)
    = -i16\pi^2\mu^{2\epsilon}\int\frac{d^Dk}{(2\pi)^D} \frac{1}{(k^2-m_{lk}^2)[(k-q-p_j)^2-m_X^2]} \nonumber\\
    &\approx \Delta+\log\left(\frac{\mu^2}{m_X^2}\right)+2
    +\left(1-\frac{m_X^2}{m_{li}^2}\right)\log\left(\frac{m_X^2}{m_X^2-m_{li}^2}\right)  \nonumber\\
    &\approx \Delta+\log\left(\frac{\mu^2}{m_X^2}\right)+1+\frac{m_{li}^2}{2m_X^2} \ ,
\end{align}
\begin{align}
C_0^X(1) &\equiv C_0(0,0,m_{li}^2,m_{lk}^2,m_X^2,m_X^2)
    =-i16\pi^2\mu^{2\epsilon}\int\frac{d^Dk}{(2\pi)^D} \frac{1}{(k^2-m_{lk}^2)[(k-p_j)^2-m_X^2][(k-q-p_j)^2-m_X^2]} \nonumber\\
    &\approx \frac{1}{2m_{li}^2}\left[2\text{Li}_2\left(\frac{m_{li}^2}{m_{li}^2-m_X^2}\right)
    +\log^2\left(\frac{m_X^2}{m_X^2-m_{li}^2}\right)\right] \nonumber\\
    &\approx -\frac{1}{m_X^2}\left(1+\frac{m_{lk}^2}{4m_X^2}\right) ,
\end{align}
\begin{align}
C_0^X(2) &\equiv C_0(0,0,m_{li}^2,m_X^2,m_{lk}^2,m_{lk}^2)
    =-i16\pi^2\mu^{2\epsilon}\int\frac{d^Dk}{(2\pi)^D} \frac{1}{(k^2-m_X^2)[(k-p_j)^2-m_{lk}^2][(k-q-p_j)^2-m_{lk}^2]} \nonumber\\
    &\approx \frac{1}{2m_{li}^2}\left[2 \text{Li}_2\left(\frac{m_{li}^2+m_{lk}^2-m_X^2}{m_{li}^2-m_X^2}\right)
    +\log^2\left(\frac{-m_{lk}^2 }{m_{li}^2-m_X^2}\right)+2 \text{Li}_2\left(1-\frac{m_X^2}{m_{lk}^2}\right) \right] \nonumber\\
    &\approx \frac{1}{m_X^2} \left[ \frac{m_{li}^2}{2m_X^2}-\frac{m_{li}^2m_{lk}^2}{2m_X^4}
    +\left(1+\frac{m_{li}^2}{2m_X^2}-\frac{m_{lk}^2}{m_X^2}\right)\log\left(\frac{m_{lk}^2}{m_X^2}\right) \right] .
\end{align}
We have crosschecked these results with the numerical software \texttt{LoopTools} \cite{Hahn:1998yk} and they are in very good agreement.
Finally, considering in (\ref{form-factor-U2}) the GIM leptonic mechanism $\sum_{k=1}^3V_{Uak}V_{Ubk}^*=0$, the leading contribution proportional to the linear mass term $m_{li}$ leads to the Eq.~(\ref{U2-loop}).

\newpage

%%%%%%%%%%%%%%%%%%%%%%%%%%%%%%%%%%%%%%%%%%%%%%%%%%%%%%%%%%%%%%%%

\newpage
%%%%%%%%%%%%%%%%%%%%%%%%%%%%%%%%%%%%%%%%%%%%%%%%%%%%%%%%%%%%%%%%%%%%%%%%%%%%%%%%%%%%%%%%%%%%%%%%%%%%%%%

%%%%%% TABLES %%%%%%

%%%%%% h --> li lj results:

\begin{table}[!h]
  \centering
\begin{tabular}{|c|c|c|c|}\hline
Deviation  & $U_{\rho 1}^h$ & $R_{\tau\tau}^\text{m331}$ & Br$(h^0\to\tau\bar{\tau})^\text{m331}$ \\
\hline
$+2\sigma$ & $0.123$         & 1.31           & 0.080 \\
$+1\sigma$ & $0.110$         & 1.05           & 0.064 \\
$0\sigma$  & $0.096$         & 0.79           & 0.048 \\
$-1\sigma$ & $0.078$         & 0.53           & 0.033 \\
$-2\sigma$ & $0.056$         & 0.27           & 0.017 \\
\hline
\end{tabular}
\caption{Comparison of the m331 model with the experimental data, $R_{\tau\tau}^\text{m331}\equiv R_{\tau\tau}^\text{Exp}=0.79\pm 0.26$ from PDG \cite{Agashe:2014kda}, Br$(h^0\to\tau\bar{\tau})^\text{SM}=0.061$. See also Fig.~\ref{FIGURE-h-tautau}.}\label{TABLE-h-tautau}
\end{table}

\begin{table}[!h]
  \centering
\begin{tabular}{|c|c|}\hline
Decay & Br$^\text{m331}$\\
\hline
$h^0\to e\bar{\mu}$    & $1.44\times10^{-14}$ \\
$h^0\to e\bar{\tau}$   & $1.92\times10^{-13}$ \\
$h^0\to \mu\bar{\tau}$ & $1.42\times10^{-14}$ \\
\hline
\end{tabular}
\caption{Lepton flavor violating tree level decays $h^0\to l_i\bar{l}_j$, with $U_{\rho1}^h=0.096$ fitted from the experimental data of $h^0\to\tau\bar{\tau}$.}\label{TABLE-h-lilj}
\end{table}

%%%%%% 3 leptons results:

\begin{table}[!h]
  \centering
\begin{tabular}{|c|c|c|}\hline
Decay & Br$^\text{Exp}$  & $\Gamma$$^\text{Exp}$ [GeV] \\
\hline
$\mu^-\to e^-e^+e^-$        & $< 10^{-12}$         & $<2.99\times10^{-31}$  \\
$\tau^-\to e^-e^+e^-$       & $< 2.7\times10^{-8}$ & $< 6.12\times10^{-20}$ \\
$\tau^-\to e^-\mu^+\mu^-$   & $< 2.7\times10^{-8}$ & $< 6.12\times10^{-20}$ \\
$\tau^-\to \mu^-e^+e^-$     & $< 1.8\times10^{-8}$ & $< 4.08\times10^{-20}$ \\
$\tau^-\to \mu^-\mu^+\mu^-$ & $< 2.1\times10^{-8}$ & $< 4.76\times10^{-20}$ \\
\hline
\end{tabular}
\caption{$l_i\to l_jl_k\bar{l}_k$ experimental upper limits from PDG \cite{Agashe:2014kda}.}\label{Table-three-leptons}
\end{table}

\begin{table}[!h]
  \centering
\begin{tabular}{|c|c|}\hline
Decay & Br$^\text{m331}$ \\
\hline
$\tau\to ee\bar{e}, \ \mu e\bar{e}      $ & $1.85\times10^{-15}$  \\
$\tau\to \mu e\bar{e},\ \mu\mu\bar{\mu} $ & $1.08\times10^{-15}$ \\
\hline
\end{tabular}
\caption{Decays $\tau\to l_jl_k\bar{l}_k$ considering $m_{U^{++}}=4.59$ TeV, the minimal bound for $m_{U^{++}}$ from $\mu\to ee\bar{e}$.}\label{TABLE-tau-3leptons}
\end{table}

%%%%%% Loops results %%%%%%

\begin{table}[!h]
  \centering
\begin{tabular}{|c|c|c|c|c|}\hline
Decay & Br$^\text{Exp}$ & $\Gamma^\text{Exp}$ [GeV] & Br$^\text{SM}$        & $\Gamma^\text{SM}$ [GeV]\\
\hline
$\mu\to e\gamma$   & $< 5.7\times10^{-13}$ & $<1.70\times10^{-31}$ & $10^{-48}$ & $10^{-67}$ \\
$\tau\to e\gamma$  & $< 3.3\times10^{-8}$  & $<7.49\times10^{-20}$ & $10^{-49}$ & $10^{-61}$ \\
$\tau\to\mu\gamma$ & $< 4.4\times10^{-8}$  & $<9.99\times10^{-20}$ & $10^{-49}$ & $10^{-61}$ \\
\hline
\end{tabular}
\caption{$l_i\to l_j\gamma$, experimental upper limits [PDG] and SM predictions.}\label{Table-loop-experiments}
\end{table}

%%%%%% Loops tables results

\begin{table}[!h]
  \centering
\begin{tabular}{|c|c|c|c|c|c|c|c|c|}\hline
\multicolumn{2}{|c|}{Scenario} & \multicolumn{7}{c|}{Br$(\mu\to e\gamma)^\text{m331}$} \\
\hline
$m_{A^0}$ [GeV] & $v_\chi O_{\rho_1}/\Lambda$ & $W$ & $U^{++}$  & $Y^{++}$ & $h^0$ & $A^0$ & Interf. & Total\\
\hline
    & 0.01 &            &                      & $3.23\times10^{-39}$ &                     &                      &
    $7.56\times10^{-34}$ & $6.84\times10^{-30}$ \\
100 & 0.1  & $10^{-48}$ & $9.61\times10^{-34}$ & $3.23\times10^{-35}$ & $1.9\times10^{-36}$ & $6.84\times10^{-30}$ & $-9.03\times10^{-33}$ & $6.83\times10^{-30}$ \\
    & 1    &            &                      & $3.23\times10^{-31}$ &                     &                      & $-9.87\times10^{-31}$ & $6.17\times10^{-30}$ \\
\hline
    & 0.01 &            &                      & $3.23\times10^{-39}$ &                     &                      & $2.25\times10^{-35}$  & $8.09\times10^{-33}$ \\
250 & 0.1  & $10^{-48}$ & $9.61\times10^{-34}$ & $3.23\times10^{-35}$ & $1.9\times10^{-36}$ & $7.11\times10^{-33}$ & $-4.97\times10^{-34}$ & $7.61\times10^{-33}$ \\
    & 1    &            &                      & $3.23\times10^{-31}$ &                     &                      & $-5.24\times10^{-32}$ & $2.79\times10^{-31}$ \\\hline
\end{tabular}
\caption{Br$(\mu\to e\gamma)^\text{m331}$ with $m_{U^{++}}=4590$ GeV and $m_{Y^{++}}=322$ GeV, see Figs.~\ref{FIGURE-loops-Sol2}(a) and (b).}\label{TABLE-muon-egamma}
\end{table}

\begin{table}[!h]
  \centering
\begin{tabular}{|c|c|c|c|c|c|c|c|c|}\hline
\multicolumn{2}{|c|}{Scenario} & \multicolumn{7}{c|}{Br$(\tau\to e\gamma)^\text{m331}$} \\
\hline
$m_{A^0}$ [GeV] & $v_\chi O_{\rho_1}/\Lambda$ & $W$ & $U^{++}$  & $Y^{++}$ & $h^0$ & $A^0$ & Interf. & Total\\
\hline
    & 0.01    &         &                      & $1.23\times10^{-32}$ &                      &                      & $-5.89\times10^{-28}$ & $2.09\times10^{-27}$ \\
100 & 0.1  & $10^{-49}$ & $1.45\times10^{-31}$ & $1.23\times10^{-28}$ & $3.36\times10^{-29}$ & $2.65\times10^{-27}$ & $1.14\times10^{-28}$ & $2.92\times10^{-27}$ \\
    & 1 &               &                      & $1.23\times10^{-24}$ &                      &                      & $7.05\times10^{-26}$ & $1.31\times10^{-24}$ \\
\hline
    & 0.01 &            &                      & $1.23\times10^{-32}$ &                      &                      & $-2.00\times10^{-29}$ & $1.66\times10^{-29}$ \\
250 & 0.1  & $10^{-49}$ & $1.45\times10^{-31}$ & $1.23\times10^{-28}$ & $3.36\times10^{-29}$ & $2.75\times10^{-30}$ & $-9.03\times10^{-29}$ & $6.97\times10^{-29}$ \\
    & 1    &            &                      & $1.23\times10^{-24}$ &                      &                      & $-7.12\times10^{-27}$ & $1.23\times10^{-24}$ \\
\hline
\end{tabular}
\caption{Br$(\tau\to e\gamma)^\text{m331}$ with $m_{U^{++}}=4590$ GeV and $m_{Y^{++}}=322$ GeV, see Figs.~\ref{FIGURE-loops-Sol2}(c) and (d).}\label{TABLE-tau-egamma}
\end{table}

\begin{table}[!h]
  \centering
\begin{tabular}{|c|c|c|c|c|c|c|c|c|}\hline
\multicolumn{2}{|c|}{Scenario} & \multicolumn{7}{c|}{Br$(\tau\to \mu\gamma)^\text{m331}$} \\
\hline
$m_{A^0}$ [GeV] & $v_\chi O_{\rho_1}/\Lambda$ & $W$ & $U^{++}$  & $Y^{++}$ & $h^0$ & $A^0$ & Interf. & Total\\
\hline
    & 0.01 &            &                      & $6.60\times10^{-33}$ &                      &                      & $-6.24\times10^{-29}$ & $4.46\times10^{-28}$ \\
100 & 0.1  & $10^{-49}$ & $7.00\times10^{-32}$ & $6.60\times10^{-29}$ & $2.07\times10^{-30}$ & $5.06\times10^{-28}$ & $1.74\times10^{-28}$ & $7.48\times10^{-28}$ \\
    & 1    &            &                      & $6.60\times10^{-25}$ &                      &                      & $2.38\times10^{-26}$ & $6.84\times10^{-25}$ \\
\hline
    & 0.01 &            &                      & $6.60\times10^{-33}$ &                      &                      & $-2.20\times10^{-30}$ & $4.75\times10^{-31}$ \\
250 & 0.1  & $10^{-49}$ & $7.00\times10^{-32}$ & $6.60\times10^{-29}$ & $2.07\times10^{-30}$ & $5.26\times10^{-31}$ & $-1.33\times10^{-29}$ & $5.53\times10^{-29}$ \\
    & 1 &               &                      & $6.60\times10^{-25}$ &                      &                      & $-1.13\times10^{-27}$ & $6.59\times10^{-25}$ \\
\hline
\end{tabular}
\caption{Br$(\tau\to\mu\gamma)^\text{m331}$ with $m_{U^{++}}=4590$ GeV and $m_{Y^{++}}=322$ GeV, see Figs.~\ref{FIGURE-loops-Sol2}(e) and (f).}\label{TABLE-tau-muongamma}
\end{table}

\newpage

%%%%%% FIGURES %%%%%%

%%%%%% h --> li lj:

%%%%%% 3 leptons:

%%%%%% Loops %%%%%%

%%%%%% Appendix figures

\begin{center}
%http://www.peteryu.ca/tutorials/publishing/latex_captions_old
\renewcommand{\thesubfigure}{\arabic{subfigure}}
\begin{figure}[!h]
\subfloat[]{\includegraphics[width=5.5cm]{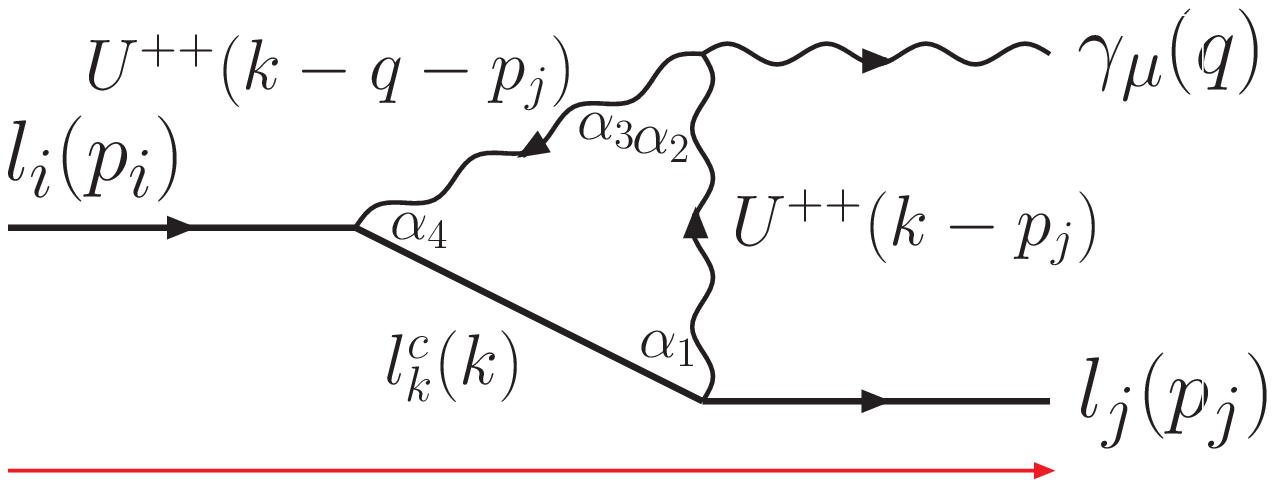}} \quad
\subfloat[]{\includegraphics[width=5.5cm]{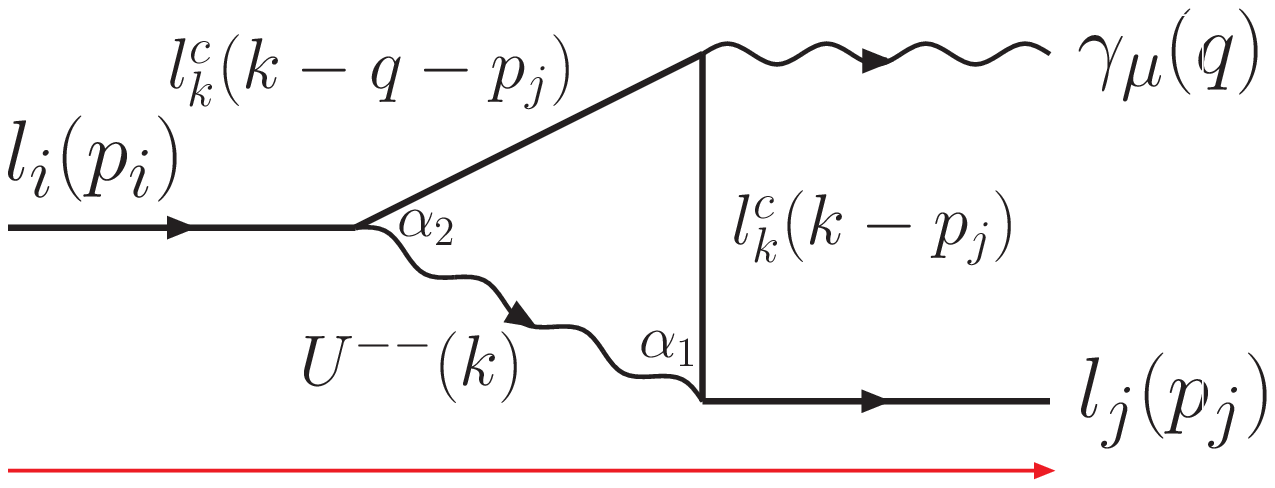}} \\
\subfloat[]{\includegraphics[width=5.5cm]{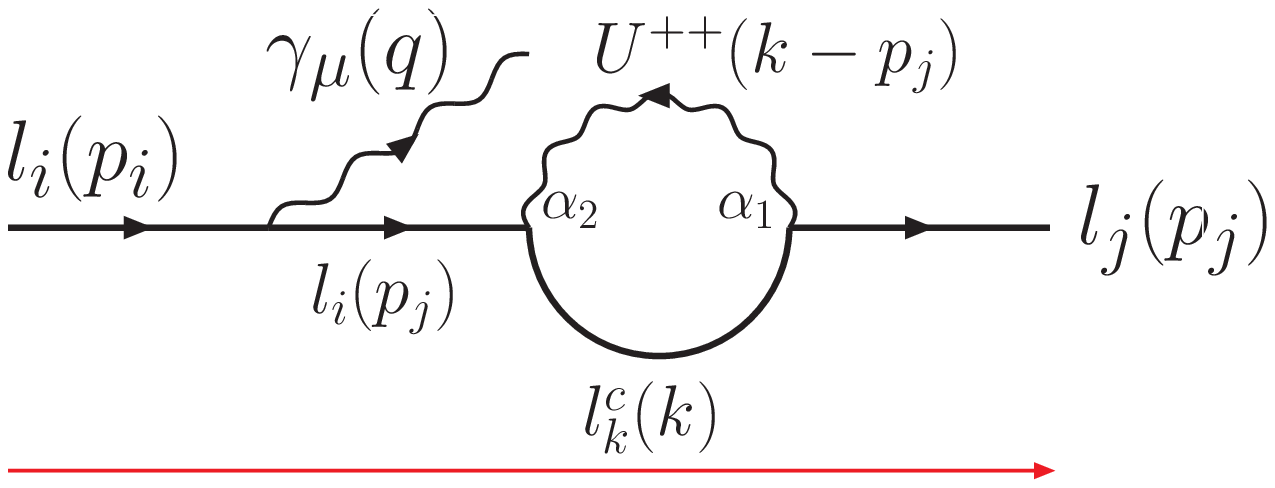}} \quad
\subfloat[]{\includegraphics[width=5.5cm]{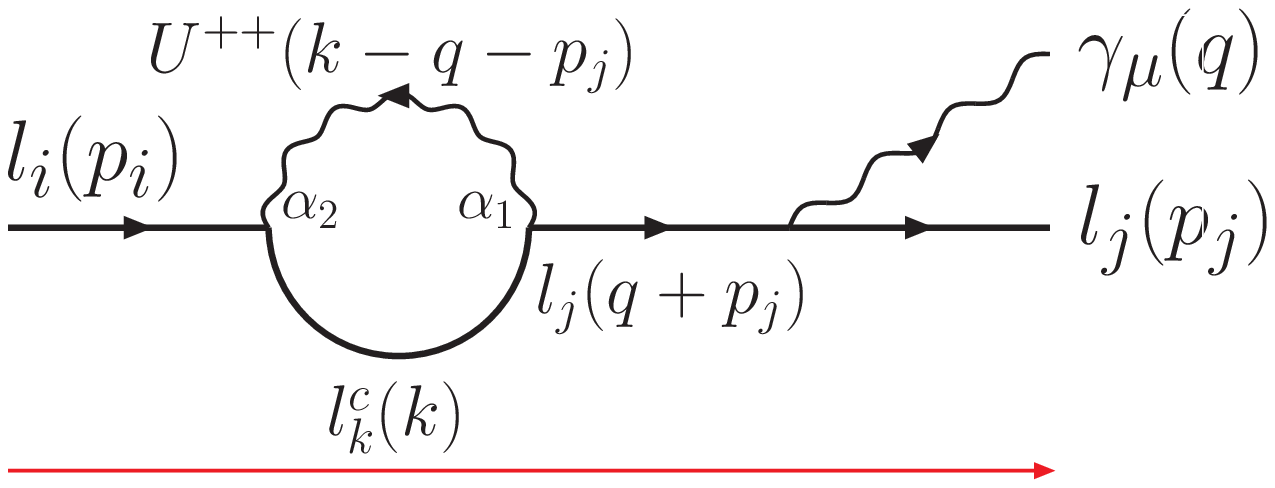}}
\caption{Decay $l_i\to l_j\gamma$ , sample of the $U^{++}\&~l_k$ contribution.}
\label{FIGURE-loops}
\end{figure}
\end{center}


\begin{thebibliography}{99}

%\cite{Boucenna:2015zwa}
\bibitem{Boucenna:2015zwa}
S.~M.~Boucenna, J.~W.~F.~Valle and A.~Vicente,
\textsl{Predicting charged lepton flavor violation from 3-3-1 gauge symmetry},
Phys.\ Rev.\ D {\bf 92}, no. 5, 053001 (2015);
%doi:10.1103/PhysRevD.92.053001
[arXiv:1502.07546 [hep-ph]].

%%%%%% PDG

%\cite{Agashe:2014kda}
\bibitem{Agashe:2014kda}
K.~A.~Olive {\it et al.} [Particle Data Group Collaboration],
\textsl{Review of Particle Physics},
Chin.\ Phys.\ C {\bf 38}, 090001 (2014) and 2015 .

\bibitem{GonzalezGarcia:2012sz}
M.~C.~Gonzalez-Garcia, M.~Maltoni, J.~Salvado and T.~Schwetz,
\textsl{Global fit to three neutrino mixing: critical look at present precision},
JHEP {\bf 1212}, 123 (2012)4;
[arXiv:1209.3023 [hep-ph]].


%\cite{Pisano:1991ee}
\bibitem{Pisano:1991ee} F. Pisano and V. Pleitez,
\textsl{$SU(3)_L\otimes U(1)_N$ model for electroweak interactions},
Phys. Rev. D \textbf{46}, 410 (1992); [hep-ph/9206242].

%\cite{Frampton:1992wt}
\bibitem{Frampton:1992wt}
P.~H.~Frampton,
\textsl{Chiral dilepton model and the flavor question},
Phys.\ Rev.\ Lett.\  {\bf 69}, 2889 (1992).

%\cite{Foot:1992rh}
\bibitem{Foot:1992rh}
R.~Foot, O.~F.~Hernandez, F.~Pisano and V.~Pleitez,
\textsl{Lepton masses in an SU(3)-L x U(1)-N gauge model},
Phys.\ Rev.\ D {\bf 47}, 4158 (1993);
[hep-ph/9207264].
%R. Foot, O. F. Hernandez, F. Pisano, and V. Pleitez, Phys. Rev. D {\bf 47}, 4158, %(1993).

\bibitem{Dias:2006ns} A.~G.~Dias, J.~C.~Montero and V.~Pleitez,
\textsl{Closing the $SU(3)_L \otimes  U(1)_X$ symmetry at electroweak scale},
Phys.\ Rev.\ D {\bf 73}, 113004 (2006); [hep-ph/0605051].

\bibitem{Machado:2013jca}
A.~C.~B.~Machado, J.~C.~Montero and V.~Pleitez,
\textsl{FCNC in the minimal 3-3-1 model revisited},
Phys.\ Rev.\ D {\bf 88}, 113002 (2013);
[arXiv:1305.1921 [hep-ph]].

%\cite{DeConto:2015eia}
\bibitem{DeConto:2015eia}
G.~De Conto, A.~C.~B.~Machado and V.~Pleitez,
\textsl{Minimal 3-3-1 model with a spectator sextet},
Phys.\ Rev.\ D {\bf 92}, no. 7, 075031 (2015);
%doi:10.1103/PhysRevD.92.075031
[arXiv:1505.01343 [hep-ph]].

\bibitem{Liu:1993gy}
J.~T.~Liu and D.~Ng,
\textsl{Lepton flavor changing processes and CP violation in the 331 model},
Phys.\ Rev.\ D {\bf 50}, 548 (1994);
[hep-ph/9401228].

%\cite{DeConto:2014fza}
\bibitem{DeConto:2014fza}
G.~De Conto and V.~Pleitez,
\textsl{Electron and neutron electric dipole moment in the 3-3-1 model with heavy leptons},
Phys.\ Rev.\ D {\bf 91}, 015006 (2015);
[arXiv:1408.6551 [hep-ph]].

%\cite{Buchmuller:1985jz}
\bibitem{Buchmuller:1985jz}
W.~Buchmuller and D.~Wyler,
\textrm{Effective Lagrangian Analysis of New Interactions and Flavor Conservation},
%doi:10.1016/0550-3213(86)90262-2

%\cite{Leung:1984ni}
\bibitem{Leung:1984ni}
C.~N.~Leung, S.~T.~Love and S.~Rao,
\textrm{Low-Energy Manifestations of a New Interaction Scale: Operator Analysis},
Z.\ Phys.\ C {\bf 31}, 433 (1986).
%doi:10.1007/BF01588041

%\cite{Chang:2005wu}
\bibitem{Chang:2005wu}
W.~F.~Chang and J.~N.~Ng,
\textrm{An Effective operators analysis of leptonic CP violation: Bridging high and low energy processes},
JHEP {\bf 0510}, 091 (2005);
%doi:10.1088/1126-6708/2005/10/091
[hep-ph/0508076].	

%\cite{Dorsner:2015mja}
\bibitem{Dorsner:2015mja}
I.~Doršner, S.~Fajfer, A.~Greljo, J.~F.~Kamenik, N.~Košnik and I.~Nišandžic,
\textrm{New Physics Models Facing Lepton Flavor Violating Higgs Decays at the Percent Level},
JHEP {\bf 1506}, 108 (2015);
%doi:10.1007/JHEP06(2015)108
[arXiv:1502.07784 [hep-ph]].

%\cite{Correia:2015tra}
\bibitem{Correia:2015tra}
F.~C.~Correia and V.~Pleitez,
\textsl{Neutral meson mixing induced by box diagrams in the 3-3-1 model with heavy leptons},
Phys.\ Rev.\ D {\bf 92}, 113006 (2015);
%doi:10.1103/PhysRevD.92.113006
[arXiv:1508.07319 [hep-ph]].

%%%%% Outras VU

%\cite{Barreto:2013paa}
\bibitem{Barreto:2013paa}
   E.~Ramirez Barreto, Y.~A.~Coutinho and J.~S.~Borges,
   \textsl{Vector-bilepton Contribution to Four Lepton Production at the LHC},
   Phys.\ Rev.\ D {\bf 88}, 035016 (2013)
%   doi:10.1103/PhysRevD.88.035016
   [arXiv:1307.4683 [hep-ph]].

%\cite{RamirezBarreto:2011av}
\bibitem{RamirezBarreto:2011av}
   E.~Ramirez Barreto, Y.~A.~Coutinho and J.~Sa Borges,
   \textsl{Vector- and Scalar-Bilepton Pair Production in Hadron Colliders},
   Phys.\ Rev.\ D {\bf 83}, 075001 (2011)
%   doi:10.1103/PhysRevD.83.075001
   [arXiv:1103.1267 [hep-ph]].


%%%%%% Experimental data h->muon tau

\bibitem{Khachatryan:2015kon}
V.~Khachatryan {\it et al.} [CMS Collaboration],
\textsl{Search for Lepton-Flavour-Violating Decays of the Higgs Boson},
Phys.\ Lett.\ B {\bf 749}, 337 (2015);
%  doi:10.1016/j.physletb.2015.07.053
[arXiv:1502.07400 [hep-ex]].

\bibitem{Aad:2015gha}
G.~Aad {\it et al.} [ATLAS Collaboration],
\textsl{Search for lepton-flavour-violating $H\to\mu\tau$ decays of the Higgs boson with the ATLAS detector},
JHEP {\bf 1511}, 211 (2015);
%  doi:10.1007/JHEP11(2015)211
[arXiv:1508.03372 [hep-ex]].

\bibitem{Chakraborty:2016gff}
I.~Chakraborty, A.~Datta and A.~Kundu,
\textsl{Lepton flavor violating Higgs boson decay $h\to\mu\tau$ at the ILC},
arXiv:1603.06681 [hep-ph].

%%%%%% More leptonic processes in 3-3-1:

%\cite{CortesMaldonado:2011uh}
\bibitem{CortesMaldonado:2011uh}
I.~Cortes Maldonado, A.~Moyotl and G.~Tavares-Velasco,
\textsl{Lepton flavor violating decay $Z\to l_i l_j$ in the 331 model},
Int.\ J.\ Mod.\ Phys.\ A {\bf 26}, 4171 (2011);
  [arXiv:1109.0661 [hep-ph]].

%\cite{Cabarcas:2013jba}
\bibitem{Cabarcas:2013jba}
J.~M.~Cabarcas, J.~Duarte and J.~-A.~Rodriguez,
\textsl{Charged lepton mixing processes in 331 Models},
Int.\ J.\ Mod.\ Phys.\ A {\bf 29}, 1450015 (2014);
[arXiv:1310.1407 [hep-ph]].

%%%%%% Vector FCNC interactions

%\cite{Aranda:2012qs}
\bibitem{Aranda:2012qs}
J.~I.~Aranda, J.~Montano, F.~Ramirez-Zavaleta, J.~J.~Toscano and E.~S.~Tututi,
\textsl{Study of the lepton flavor-violating $Z'\to\tau\mu$ decay},
Phys.\ Rev.\ D {\bf 86}, 035008 (2012);
%  doi:10.1103/PhysRevD.86.035008
[arXiv:1202.6288 [hep-ph]].

%\cite{Barger:1987nn}
\bibitem{Barger:1987nn}
V.~D.~Barger and R.~J.~N.~Phillips,
\textsl{Collider Physics, Updated Edition},
%REDWOOD CITY, USA:
ADDISON-WESLEY (1996) 592 P. (FRONTIERS IN PHYSICS, 71)

%%%%%% W loop

\bibitem{W-loop}
%\cite{Petcov:1976ff}
%\bibitem{Petcov:1976ff}
S.~T.~Petcov,
\textsl{The Processes $\mu\to e \gamma, \mu\to ee\bar{e}, \nu'\to\nu\gamma$ in the Weinberg-Salam Model with Neutrino Mixing},
Sov.\ J.\ Nucl.\ Phys.\  {\bf 25}, 340 (1977)
[Yad.\ Fiz.\  {\bf 25}, 641 (1977)]
Erratum: [Sov.\ J.\ Nucl.\ Phys.\  {\bf 25}, 698 (1977)]
Erratum: [Yad.\ Fiz.\  {\bf 25}, 1336 (1977)].
  %\cite{Cheng:1976uq}
%\bibitem{Cheng:1976uq}
T.~P.~Cheng and L.~F.~Li,
\textsl{Nonconservation of Separate $\mu$-Lepton and $e$-Lepton Numbers in Gauge Theories with $v+a$ Currents},
Phys.\ Rev.\ Lett.\  {\bf 38}, 381 (1977).
%  doi:10.1103/PhysRevLett.38.381
%\cite{Lee:1977tib}
%\bibitem{Lee:1977tib}
B.~W.~Lee and R.~E.~Shrock,
\textsl{Natural Suppression of Symmetry Violation in Gauge Theories: Muon - Lepton and Electron Lepton Number Nonconservation},
Phys.\ Rev.\ D {\bf 16}, 1444 (1977).
%  doi:10.1103/PhysRevD.16.1444
%\cite{Marciano:1977wx}
%\bibitem{Marciano:1977wx}
W.~J.~Marciano and A.~I.~Sanda,
\textsl{Exotic Decays of the Muon and Heavy Leptons in Gauge Theories},
Phys.\ Lett.\ B {\bf 67}, 303 (1977).
%  doi:10.1016/0370-2693(77)90377-X
%\cite{Altarelli:1977zq}
%\bibitem{Altarelli:1977zq}
G.~Altarelli, L.~Baulieu, N.~Cabibbo, L.~Maiani and R.~Petronzio,
\textsl{Muon Number Nonconserving Processes in Gauge Theories of Weak Interactions},
Nucl.\ Phys.\ B {\bf 125}, 285 (1977).
Erratum: [Nucl.\ Phys.\ B {\bf 130}, 516 (1977)].
%doi:10.1016/0550-3213(77)90407-2, 10.1016/0550-3213(77)90254-1

\bibitem{Cheng-Lee-book}
T. P. Cheng and L. F. Li, \textsl{Gauge Theory of Elementary Particle Physics}, Oxford University Press, New York (1984).

%\cite{Chatrchyan:2013wfa}
\bibitem{Chatrchyan:2013wfa}
S.~Chatrchyan {\it et al.} [CMS Collaboration],
\textrm{Search for top-quark partners with charge 5/3 in the same-sign
dilepton final state},
Phys.\ Rev.\ Lett.\  {\bf 112}, no. 17, 171801 (2014);
%doi:10.1103/PhysRevLett.112.171801
[arXiv:1312.2391 [hep-ex]].

%\cite{Montero:1992jk}
\bibitem{Montero:1992jk}
J.~C.~Montero, F.~Pisano and V.~Pleitez,
\textsl{Neutral currents and GIM mechanism in $SU(3)_L \times U(1)_N$ models for electroweak interactions},
Phys.\ Rev.\ D {\bf 47}, 2918 (1993);
%doi:10.1103/PhysRevD.47.2918
[hep-ph/9212271].

%\cite{Dong:2008sw}
\bibitem{Dong:2008sw}
P.~V.~Dong and H.~N.~Long,
\textsl{Neutrino masses and lepton flavor violation in the 3-3-1 model with right-handed neutrinos},
Phys.\ Rev.\ D {\bf 77}, 057302 (2008);
%doi:10.1103/PhysRevD.77.057302
[arXiv:0801.4196 [hep-ph]].

%\cite{Hernandez:2015tna}
\bibitem{Hernandez:2015tna}
A.~E.~Cárcamo Hernández and R.~Martinez,
\textsl{A predictive 3-3-1 model with $A_4$ flavor symmetry},
Nucl.\ Phys.\ B {\bf 905}, 337 (2016)
%  doi:10.1016/j.nuclphysb.2016.02.025
% [arXiv:1501.05937 [hep-ph]].

%\cite{Auger:2012ar}
\bibitem{Auger:2012ar}
M.~Auger {\it et al.} [EXO-200 Collaboration],
\textsl{Search for Neutrinoless Double-Beta Decay in $^{136}$Xe with EXO-200},
Phys.\ Rev.\ Lett.\  {\bf 109}, 032505 (2012)
%doi:10.1103/PhysRevLett.109.032505
[arXiv:1205.5608 [hep-ex]].

%\cite{Mohapatra:1981pm}
\bibitem{Mohapatra:1981pm}
R.~N.~Mohapatra and J.~D.~Vergados,
\textsl{A New Contribution to Neutrinoless Double Beta Decay in Gauge Models},
Phys.\ Rev.\ Lett.\  {\bf 47}, 1713 (1981).
%doi:10.1103/PhysRevLett.47.1713

%\cite{Schechter:1981bd}
\bibitem{Schechter:1981bd}
J.~Schechter and J.~W.~F.~Valle,
\textsl{Neutrinoless Double beta Decay in SU(2) x U(1) Theories},
Phys.\ Rev.\ D {\bf 25}, 2951 (1982).
%doi:10.1103/PhysRevD.25.2951

%\cite{Wolfenstein:1982bf}
\bibitem{Wolfenstein:1982bf}
L.~Wolfenstein,
\textsl{Triplet Scalar Bosons and Double Beta Decay},
Phys.\ Rev.\ D {\bf 26}, 2507 (1982).
%doi:10.1103/PhysRevD.26.2507

%\cite{Escobar:1982ec}
\bibitem{Escobar:1982ec}
C.~O.~Escobar and V.~Pleitez,
\textsl{Some New Contributions to Neutrinoless Double Beta Decay in an SU(2) X U(1) Model},
Phys.\ Rev.\ D {\bf 28}, 1166 (1983).
%doi:10.1103/PhysRevD.28.1166

%\cite{Marciano:2008zz}
\bibitem{Marciano:2008zz}
   W.~J.~Marciano, T.~Mori and J.~M.~Roney,
\textsl{Charged Lepton Flavor Violation Experiments},
   Ann.\ Rev.\ Nucl.\ Part.\ Sci.\  {\bf 58}, 315 (2008).
   %doi:10.1146/annurev.nucl.58.110707.171126

%\cite{Wu:2016gjv}
\bibitem{Wu:2016gjv}
   C.~Wu,
\textsl{Search for Muon to electron conversion at J-PARC},
   Hyperfine Interact.\  {\bf 237}, no. 1, 149 (2016).
   %doi:10.1007/s10751-016-1351-0

%\cite{Pleitez:1999ix}
\bibitem{Pleitez:1999ix}
   V.~Pleitez,
\textsl{A Remark on the muonium to anti-muonium conversion in a 331 model},
   Phys.\ Rev.\ D {\bf 61}, 057903 (2000)
   %doi:10.1103/PhysRevD.61.057903
   [hep-ph/9905406].

%\cite{Bernstein:2013hba}
\bibitem{Bernstein:2013hba}
   R.~H.~Bernstein and P.~S.~Cooper,
\textsl{Charged Lepton Flavor Violation: An Experimenter's Guide},
   Phys.\ Rept.\  {\bf 532}, 27 (2013)
   %doi:10.1016/j.physrep.2013.07.002
   [arXiv:1307.5787 [hep-ex]].

%\cite{Willmann:1998gd}
\bibitem{Willmann:1998gd}
   L.~Willmann {\it et al.},
\textrm{New bounds from searching for muonium to anti-muonium
conversion},
   Phys.\ Rev.\ Lett.\  {\bf 82}, 49 (1999);
%doi:10.1103/PhysRevLett.82.49
   [hep-ex/9807011].

%\cite{Denner:1992vza}
\bibitem{Denner:1992vza}
A.~Denner, H.~Eck, O.~Hahn and J.~Kublbeck,
\textsl{Feynman rules for fermion number violating interactions},
Nucl.\ Phys.\ B {\bf 387}, 467 (1992).

%\cite{Denner:1992me}
\bibitem{Denner:1992me}
A.~Denner, H.~Eck, O.~Hahn and J.~Kublbeck,
\textsl{Compact Feynman rules for Majorana fermions},
Phys.\ Lett.\ B {\bf 291}, 278 (1992).

%%%%%% Old majorana treatment:

%\cite{Jones:1983eh}
\bibitem{Jones:1983eh}
S.~K.~Jones and C.~H.~Llewellyn Smith,
\textsl{Leptoproduction of Supersymmetric Particles},
Nucl.\ Phys.\ B {\bf 217}, 145 (1983).
%  doi:10.1016/0550-3213(83)90082-2

%\cite{Haber:1984rc}
\bibitem{Haber:1984rc}
H.~E.~Haber and G.~L.~Kane,
\textsl{The Search for Supersymmetry: Probing Physics Beyond the Standard Model},
Phys.\ Rept.\  {\bf 117}, 75 (1985).
%  doi:10.1016/0370-1573(85)90051-1

%\cite{Gates:1987ay}
\bibitem{Gates:1987ay}
E.~I.~Gates and K.~L.~Kowalski,
\textsl{Majorana Feynman Rules},
Phys.\ Rev.\ D {\bf 37}, 938 (1988).
%  doi:10.1103/PhysRevD.37.938

%\cite{Gluza:1991wj}
\bibitem{Gluza:1991wj}
J.~Gluza and M.~Zralek,
\textsl{Feynman rules for Majorana neutrino interactions},
Phys.\ Rev.\ D {\bf 45}, 1693 (1992).
%  doi:10.1103/PhysRevD.45.1693

%\cite{Bu:2008fx}
\bibitem{Bu:2008fx}
J.~P.~Bu, Y.~Liao and J.~Y.~Liu,
\textsl{Lepton Flavor Violating Muon Decays in a Model of Electroweak-Scale Right-Handed Neutrinos},
Phys.\ Lett.\ B {\bf 665}, 39 (2008);
%doi:10.1016/j.physletb.2008.05.059
[arXiv:0802.3241 [hep-ph]].


%%%%%% FeynCal, Package-X

%\cite{Mertig:1990an}
\bibitem{Mertig:1990an}
R.~Mertig, M.~Bohm and A.~Denner,
\textsl{FEYN CALC: Computer algebraic calculation of Feynman amplitudes},
Comput.\ Phys.\ Commun.\  {\bf 64}, 345 (1991).
%  doi:10.1016/0010-4655(91)90130-D

%\cite{Shtabovenko:2016sxi}
\bibitem{Shtabovenko:2016sxi}
V.~Shtabovenko, R.~Mertig and F.~Orellana,
\textsl{New Developments in FeynCalc 9.0},
arXiv:1601.01167 [hep-ph].

\bibitem{Lavoura:2003xp}
  L.~Lavoura,
  \textsl{General formulae for $f(1) \to f(2)\gamma$},
  Eur.\ Phys.\ J.\ C {\bf 29}, 191 (2003)
  doi:10.1140/epjc/s2003-01212-7
  [hep-ph/0302221].
  %%CITATION = doi:10.1140/epjc/s2003-01212-7;%%
  %71 citations counted in INSPIRE as of 30 Sep 2016

\bibitem{Romao-book}
Jorge C. Rom\~ao,
\textsl{Modern Techniques for One-Loop Calculations}, (2006).
http://porthos.ist.utl.pt/OneLoop/one-loop.pdf

%%%%%%%%%% fi->fj\gamma considering bubbles:

\bibitem{references-with-bubbles}

%\cite{Leontaris:1985qc}
%\bibitem{Leontaris:1985qc}
  G.~K.~Leontaris, K.~Tamvakis and J.~D.~Vergados,
  \textsl{Lepton and Family Number Violation From Exotic Scalars},
  Phys.\ Lett.\ B {\bf 162}, 153 (1985).
  %doi:10.1016/0370-2693(85)91078-0
  %%CITATION = doi:10.1016/0370-2693(85)91078-0;%%
  %34 citations counted in INSPIRE as of 17 Oct 2016
%\cite{Vergados:1985pq}
%\bibitem{Vergados:1985pq}
  J.~D.~Vergados,
  \textsl{The Neutrino Mass and Family, Lepton and Baryon Nonconservation in Gauge Theories},
  Phys.\ Rept.\  {\bf 133}, 1 (1986).
  %doi:10.1016/0370-1573(86)90088-8
  %%CITATION = doi:10.1016/0370-1573(86)90088-8;%%
  %265 citations counted in INSPIRE as of 17 Oct 2016
%\cite{Kosmas:1993ch}
%\bibitem{Kosmas:1993ch}
  T.~S.~Kosmas, G.~K.~Leontaris and J.~D.~Vergados,
  \textsl{Lepton flavor nonconservation},
  Prog.\ Part.\ Nucl.\ Phys.\  {\bf 33}, 397 (1994)
  %doi:10.1016/0146-6410(94)90047-7
  [hep-ph/9312217].
  %%CITATION = doi:10.1016/0146-6410(94)90047-7;%%
  %60 citations counted in INSPIRE as of 17 Oct 2016
%\cite{Cortes-Maldonado:2013rca}
%\bibitem{Cortes-Maldonado:2013rca}
  I.~Cortés-Maldonado, G.~Hernández-Tomé and G.~Tavares-Velasco,
  \textsl{Decay $t\to c\gamma$ in models with $SU_L(3)\times U_X(1)$ gauge symmetry},
  Phys.\ Rev.\ D {\bf 88}, no. 1, 014011 (2013)
  %doi:10.1103/PhysRevD.88.014011
  [arXiv:1305.2606 [hep-ph]].
  %%CITATION = doi:10.1103/PhysRevD.88.014011;%%
  %2 citations counted in INSPIRE as of 17 Oct 2016
%\cite{Ilakovac:2012sh}
%\bibitem{Ilakovac:2012sh}
  A.~Ilakovac, A.~Pilaftsis and L.~Popov,
  \textsl{Charged lepton flavor violation in supersymmetric low-scale seesaw models},
  Phys.\ Rev.\ D {\bf 87}, no. 5, 053014 (2013)
  %doi:10.1103/PhysRevD.87.053014
  [arXiv:1212.5939 [hep-ph]].
  %%CITATION = doi:10.1103/PhysRevD.87.053014;%%
  %38 citations counted in INSPIRE as of 17 Oct 2016
%\cite{Abada:2014kba}
%\bibitem{Abada:2014kba}
  A.~Abada, M.~E.~Krauss, W.~Porod, F.~Staub, A.~Vicente and C.~Weiland,
  \textsl{Lepton flavor violation in low-scale seesaw models: SUSY and non-SUSY contributions},
  JHEP {\bf 1411}, 048 (2014)
  %doi:10.1007/JHEP11(2014)048
  [arXiv:1408.0138 [hep-ph]].
  %%CITATION = doi:10.1007/JHEP11(2014)048;%%
  %36 citations counted in INSPIRE as of 17 Oct 2016

%%%%%%%%%%%%%%%%%%%%%%%%%%%%%%%%%%%%%%%%%%%%%%%%%%%%%

%\cite{Montano:2005gs}
\bibitem{Montano:2005gs}
J.~Montano, G.~Tavares-Velasco, J.~J.~Toscano and F.~Ramirez-Zavaleta,
\textsl{$SU_L(2) \times U_Y(1)$-invariant description of the bilepton contribution to the $WWV$ vertex in the minimal 331 model},
Phys.\ Rev.\ D {\bf 72}, 055023 (2005);
%  doi:10.1103/PhysRevD.72.055023
[hep-ph/0508166].
%%CITATION = doi:10.1103/PhysRevD.72.055023;%%
%22 citations counted in INSPIRE as of 27 Jun 2016

%\cite{Patel:2015tea}
\bibitem{Patel:2015tea}
H.~H.~Patel,
\textsl{Package-X: A Mathematica package for the analytic calculation of one-loop integrals},
Comput.\ Phys.\ Commun.\  {\bf 197}, 276 (2015);
%  doi:10.1016/j.cpc.2015.08.017
[arXiv:1503.01469 [hep-ph]].

\bibitem{Hahn:1998yk}
  T.~Hahn and M.~Perez-Victoria,
  \textsl{Automatized one loop calculations in four-dimensions and D-dimensions},
  Comput.\ Phys.\ Commun.\  {\bf 118}, 153 (1999)
%  doi:10.1016/S0010-4655(98)00173-8
  [hep-ph/9807565].
  %%CITATION = doi:10.1016/S0010-4655(98)00173-8;%%
  %1154 citations counted in INSPIRE as of 14 Jul 2016

\end{thebibliography}
\end{document}